\documentclass[twocolumn,trackchanges]{aastex62}

\shorttitle{WINERED line catalog: A-type star}
\shortauthors{Sameshima et al.}

\usepackage{booktabs}

\begin{document}

\title{WINERED high-resolution near-infrared line catalog: A-type star}

\email{sameshima@cc.kyoto-su.ac.jp}

\author{Hiroaki Sameshima}
\affiliation{Laboratory of Infrared High-resolution spectroscopy (LiH), Koyama
Astronomical Observatory, Kyoto Sangyo University, Motoyama, Kamigamo,
Kita-ku, Kyoto 603-8555, Japan}

\author{Yuji Ikeda}
\affiliation{Laboratory of Infrared High-resolution spectroscopy (LiH), Koyama
Astronomical Observatory, Kyoto Sangyo University, Motoyama, Kamigamo,
Kita-ku, Kyoto 603-8555, Japan}
\affiliation{Photocoding, 460-102 Iwakura-Nakamachi, Sakyo-ku, Kyoto,606-0025, Japan}

\author{Noriyuki Matsunaga}
\affiliation{Department of Astronomy, Graduate School of Science, The
University of Tokyo, 7-3-1 Hongo, Bunkyo-ku, Tokyo 113-0033, Japan}
\affiliation{Laboratory of Infrared High-resolution spectroscopy (LiH), Koyama
Astronomical Observatory, Kyoto Sangyo University, Motoyama, Kamigamo,
Kita-ku, Kyoto 603-8555, Japan}

\author{Kei Fukue}
\affiliation{Laboratory of Infrared High-resolution spectroscopy (LiH), Koyama
Astronomical Observatory, Kyoto Sangyo University, Motoyama, Kamigamo,
Kita-ku, Kyoto 603-8555, Japan}

\author{Naoto Kobayashi}
\affiliation{Kiso Observatory, Institute of Astronomy, School of
Science, The University of Tokyo, 10762-30 Mitake, Kiso-machi, Kiso-gun,
Nagano, 397-0101, Japan}
\affiliation{Institute of Astronomy, School of Science, The University of
Tokyo, 2-21-1 Osawa, Mitaka, Tokyo 181-0015, Japan}
\affiliation{Laboratory of Infrared High-resolution spectroscopy (LiH), Koyama
Astronomical Observatory, Kyoto Sangyo University, Motoyama, Kamigamo,
Kita-ku, Kyoto 603-8555, Japan}

\author{Sohei Kondo}
\affiliation{Laboratory of Infrared High-resolution spectroscopy (LiH), Koyama
Astronomical Observatory, Kyoto Sangyo University, Motoyama, Kamigamo,
Kita-ku, Kyoto 603-8555, Japan}

\author{Satoshi Hamano}
\affiliation{Laboratory of Infrared High-resolution spectroscopy (LiH), Koyama
Astronomical Observatory, Kyoto Sangyo University, Motoyama, Kamigamo,
Kita-ku, Kyoto 603-8555, Japan}

\author{Hideyo Kawakita}
\affiliation{Laboratory of Infrared High-resolution spectroscopy (LiH), Koyama
Astronomical Observatory, Kyoto Sangyo University, Motoyama, Kamigamo,
Kita-ku, Kyoto 603-8555, Japan}
\affiliation{Department of Physics, Faculty of Sciences, Kyoto Sangyo
University, Motoyama, Kamigamo, Kita-ku, Kyoto 603-8555, Japan}

\author{Chikako Yasui}
\affiliation{National Astronomical Observatory of Japan, 2-21-1 Osawa,
Mitaka, Tokyo 181-8588, Japan}
\affiliation{Laboratory of Infrared High-resolution spectroscopy (LiH), Koyama
Astronomical Observatory, Kyoto Sangyo University, Motoyama, Kamigamo,
Kita-ku, Kyoto 603-8555, Japan}

\author{Natsuko Izumi}
\affiliation{National Astronomical Observatory of Japan, 2-21-1 Osawa,
Mitaka, Tokyo 181-8588, Japan}

\author{Misaki Mizumoto}
\affiliation{Centre for Extragalactic Astronomy, Department of Physics,
University of Durham, South Road, Durham DH1 3LE, UK}

\author{Shogo Otsubo}
\affiliation{Laboratory of Infrared High-resolution spectroscopy (LiH), Koyama
Astronomical Observatory, Kyoto Sangyo University, Motoyama, Kamigamo,
Kita-ku, Kyoto 603-8555, Japan}
\affiliation{Department of Physics, Faculty of Sciences, Kyoto Sangyo
University, Motoyama, Kamigamo, Kita-ku, Kyoto 603-8555, Japan}

\author{Keiichi Takenaka}
\affiliation{Laboratory of Infrared High-resolution spectroscopy (LiH), Koyama
Astronomical Observatory, Kyoto Sangyo University, Motoyama, Kamigamo,
Kita-ku, Kyoto 603-8555, Japan}
\affiliation{Department of Physics, Faculty of Sciences, Kyoto Sangyo
University, Motoyama, Kamigamo, Kita-ku, Kyoto 603-8555, Japan}

\author{Ayaka Watase}
\affiliation{Laboratory of Infrared High-resolution spectroscopy (LiH), Koyama
Astronomical Observatory, Kyoto Sangyo University, Motoyama, Kamigamo,
Kita-ku, Kyoto 603-8555, Japan}
\affiliation{Department of Physics, Faculty of Sciences, Kyoto Sangyo
University, Motoyama, Kamigamo, Kita-ku, Kyoto 603-8555, Japan}

\author{Akira Asano}
\affiliation{Laboratory of Infrared High-resolution spectroscopy (LiH), Koyama
Astronomical Observatory, Kyoto Sangyo University, Motoyama, Kamigamo,
Kita-ku, Kyoto 603-8555, Japan}
\affiliation{Department of Physics, Faculty of Sciences, Kyoto Sangyo
University, Motoyama, Kamigamo, Kita-ku, Kyoto 603-8555, Japan}

\author{Tomohiro Yoshikawa}
\affiliation{Edechs, 17-203 Iwakura-Minami-Osagi-cho, Sakyo-ku, Kyoto
606-0003, Japan}

%%%%%%%%%%%%%%%%%%%%%%%%%%%%%%%%%%%%%%%%%%
% Abstract
%%%%%%%%%%%%%%%%%%%%%%%%%%%%%%%%%%%%%%%%%%
 \begin{abstract}
  We present a catalog of absorption lines in the $z^\prime, Y,$ and $J$
  bands that we identified in 21 Lyn, a slowly rotating A0.5\,V star.
  We detected 155 absorption features in the high-resolution
  (0.90--1.35~\micron, $R = 28,000$) spectrum obtained with the WINERED
  spectrograph after the telluric absorption was carefully removed using
  a spectrum of a B-type star as a telluric standard.  With a visual
  comparison with synthetic spectra, we compiled a catalog of 219 atomic
  lines for the 155 features, some of which are composed of multiple
  fine structure lines.  The high-quality WINERED spectrum enabled us to
  detect a large number of weak lines down to $\sim$1\% in depth, which
  are identified for an A-type star for the first time.  The catalog
  includes the lines of H, C, N, O, Mg, Al, Si, S, Ca, Fe, and Sr.
  These new lines are expected to be useful for spectral classification
  and chemical abundance analyses, whilst the line list is useful for
  observers who plan to use A-type stars as telluric standards because
  it is necessary to distinguish between stellar lines and telluric
  absorption lines in high-resolution spectra.  ASCII versions of the
  spectra are available in the online version of the journal.
 \end{abstract}


%%%%%%%%%%%%%%%%%%%%%%%%%%%%%%%%%%%%%%%%%%
% Introduction
%%%%%%%%%%%%%%%%%%%%%%%%%%%%%%%%%%%%%%%%%%
\section{Introduction}

Recent progress in near-infrared (NIR) spectrographs has enabled us to
obtain NIR high-resolution and high-quality spectra of stars and opened
the possibility of deriving fundamental parameters including chemical
abundances.  However, reliable line catalogs in the NIR region have yet
to be established.  One of the most reliable NIR line catalogs based on
astronomical observation was given by \cite{1999ApJS..124..527M}; they
built a line catalog in the wavelength ranges of 1.00--1.34 and
1.49--1.80 \micron\ using the solar spectrum obtained with a Fourier
transform spectrometer (\citealt{1991aass.book.....L}).  However,
because their line catalog is based on the solar spectrum, it does not
cover lines that do not appear in the solar spectrum and hence is
applicable only to stars with limited spectral types.

Aiming at extending the coverage of line catalogs in both the wavelength
and spectral type directions, we are carrying out a project to establish
NIR line catalogs based on observations of various types of stars using
our high-dispersion echelle spectrograph, WINERED
(\citealt{2016SPIE.9908E..5ZI}).  In the WIDE mode, WINERED covers the
wavelength range of 0.90--1.35~\micron\ (corresponding to the $z^\prime,
Y$, and $J$ bands) with a resolving power of $R \equiv
\lambda/\Delta\lambda = 28,000$.  The high throughput of WINERED enables
us to obtain high-quality spectra---the typical signal-to-noise ratio
being $\gtrsim 300$ for luminous stars---with short-time exposures.

As the first step of our project, we present a line catalog produced
from the spectra of an A-type star obtained with WINERED.  Surprisingly,
very little spectroscopic study has been carried out for A-type stars in
the wavelength range of 0.90--1.35~\micron.  The most reliable spectral
atlas of A-type stars in this range was previously given by
\cite{2000ApJ...535..325W}, who presented $J$-band spectra for 88
fundamental MK standard stars.  However, because the resolving power was
limited to $R \sim 3000$, their spectra of A-type stars such as HR 4534
(A3\,V) show a very limited number of lines besides hydrogen lines.
High-resolution spectroscopic observation is thus necessary to
investigate the weak absorption lines in A-type stars.  On the other
hand, given that the spectral lines in A-type stars are rotationally
broadened in general, the resolution of WINERED is sufficient to resolve
each line.

The line catalog of an A-type star is useful not only for scientific
studies of A-type stars but also for making use of A-type stars as
telluric standards.  In low-resolution spectroscopic observations,
A-type stars have been widely used as telluric standard stars owing to
their relatively featureless spectra (except for strong hydrogen lines).
However, their weak metal lines get resolved when the spectral
resolution is high (e.g., $R > 10,000$), which complicates the telluric
correction unless the intrinsic stellar lines are carefully removed
(\citealt{2018PASP..130g4502S}).  The line catalog in the present work
would help to distinguish between stellar lines and telluric absorption
lines in the spectra of A-type stars.

Throughout this paper, we use air wavelengths rather than vacuum
wavelengths unless otherwise noted.

\section{Data acquisition and reduction}

\begin{deluxetable}{lcc}[t]
\tablecaption{Observation Log \label{tab:obslog}}
\tablehead{
\colhead{Parameter} & \colhead{21 Lyn} & \colhead{HD 43384} 
 }
 \startdata
 Spectral type  & A0.5\,V & B3\,Iab \\
 Obs. Time (UT) & 2014 Jan 23 15:00 & 2014 Jan 23 14:26 \\
 Airmass       & 1.04--1.06 & 1.07--1.10 \\
 Exposure      & 200 s $\times\ 6$ & 360 s $\times\ 4$ \\
 Dithering     & ABBAAB & ABBA \\
 Signal-to-noise ratio & 830 & 580 \\
 \enddata
 \tablecomments{The spectral types are retrieved from the SIMBAD
 astronomical database.  The signal-to-noise ratio is measured from
 the standard deviation of the continuum level of the coadded spectrum at
 $\sim$1.04~\micron, where telluric absorption is almost negligible.}
\end{deluxetable}

A slowly rotating star is desirable as our target to produce the line
catalog because the absorption lines are relatively deep and less
affected by line blending and the identification task becomes easier.
However, slowly rotating A-type stars often show chemical peculiarities
(e.g., \citealt{1974ARA&A..12..257P,1995ApJS...99..135A}).
\cite{2014AA...562A..84R} performed a cluster analysis of 47 A0--A1
stars with low projected rotational velocity (hereafter, $v\sin i$) and
split them into chemically peculiar (CP) and normal stars.  Among the
normal stars that they identified, 21 Lyn (A0.5\,V, $v \sin
i=19~\mathrm{km~s^{-1}}$) was observed with WINERED, giving a
high-quality spectrum that can be used for our purpose.

The observation of 21 Lyn was carried out on January 23, 2014 using the
WINERED echelle spectrograph mounted on the 1.3-m Araki Telescope at the
Koyama Astronomical Observatory in Kyoto, Japan.  The observation mode
was set to the WIDE mode with the 100-\micron-width slit, which realizes
a coverage of 0.90--1.35~\micron\ and a resolving power of $R=28,000$.
The target was observed at two positions separated by about 30\arcsec\
along the slit by nodding the telescope to make the ABBA dithering
sequence.  The observation log is summarized in Table \ref{tab:obslog}.

All data were reduced in a standard manner using IRAF\footnote{ IRAF is
distributed by the National Optical Astronomy Observatories, which are
operated by the Association of Universities for Research in Astronomy,
Inc., under cooperative agreement with the National Science Foundation.}
routines as follows.  Sky subtraction was performed by taking the
difference between two consecutive images taken at different slit
positions, i.e., A$-$B and B$-$A.  Scattered light was evaluated at the
interorder regions of each difference image and then removed.  Flat
fielding was performed using a dome-flat image.  Bad pixels were then
masked and replaced by linear interpolation from the surrounding pixels.
Owing to the large value of the $\gamma$ angle of WINERED, the spectral
lines in the two-dimensional images were tilted with respect to the
dispersion direction; this tilt was corrected by performing a
geometrical transformation using arc-lamp images as a reference.  Then,
one-dimensional spectra were extracted using the IRAF task {\tt apall}.
Wavelength calibration was performed using Th--Ar lamp spectra that were
extracted in the same way as the target object.  Normalization of each
frame was performed by the IRAF task {\tt continuum}, where cubic spline
curves or low-order Legendre polynomials were mainly used to fit the
continuum.  These frames were then coadded by averaging the counts for
each pixel, in which spurious features were carefully checked by eye and
masked.

The B-type star HD 43384 (B3\,Iab), which was observed just before 21
Lyn with almost the same airmass (see Table \ref{tab:obslog}), was used
as a telluric calibration source.  From the continuum-normalized
spectrum of HD 43384 reduced in the same way as 21 Lyn, intrinsic
stellar lines were carefully distinguished from telluric absorption
lines by using synthetic telluric spectra created by {\tt molecfit}
(\citealt{2015A&A...576A..77S,2015A&A...576A..78K}) as a reference.
These stellar lines and features other than telluric absorption were
removed by fitting multiple Gaussian curves (see
\citealt{2018PASP..130g4502S} for the details).  The telluric spectra
retrieved in this way were used to remove the telluric absorption from
the 21 Lyn spectra by the IRAF task {\tt telluric}, where the difference
in the effective airmass was corrected following Beer's law
(\citealt{1852AnP...162...78B}).  Finally, we obtained the telluric
corrected spectrum of 21 Lyn for the wavelength ranges of 0.910--0.930,
0.960--1.115, and 1.160--1.330~\micron, which correspond to the
$z^\prime, Y$, and $J$ bands, respectively.  Note that we could not
obtain the appropriate spectra for the wavelength range of
1.307--1.312~\micron\ owing to the nonlinear response of the bad pixel
region on the array; we decided not to use this part of the spectra in
the following analysis.

After telluric correction, continuum normalization was again performed
by the IRAF task {\tt continuum} to improve the normalization.  This was
especially important around the wavelength ranges where the first
normalization performed before telluric correction was complicated by
telluric absorption lines.  Note that we could not perform normalization
around the hydrogen lines in the straightforward manner described above
because these lines were often not fully covered within a single order
owing to their large width.  We therefore determined the continuum
levels around the hydrogen lines so that their line profiles match
synthetic spectra, which prevents quantitative discussions about the
hydrogen lines.  The continuum-normalized spectra of 21 Lyn are shown in
Figure \ref{fig:spectra}, where the wavelengths are not heliocentric but
corrected to be at rest in air by cross-correlation matching with
synthetic spectra.  In the figure, the telluric spectra created above
are also wavelength-shifted by the same amount of 21 Lyn and shown in
the lower panels.  The symbol $\oplus$ indicates the spectral regions
where we concluded that spurious features remain even after the telluric
correction.  ASCII versions of the 21 Lyn and the telluric spectra are
available in the online version of the journal.

\begin{figure*}[p]
 \epsscale{1.1} \plotone{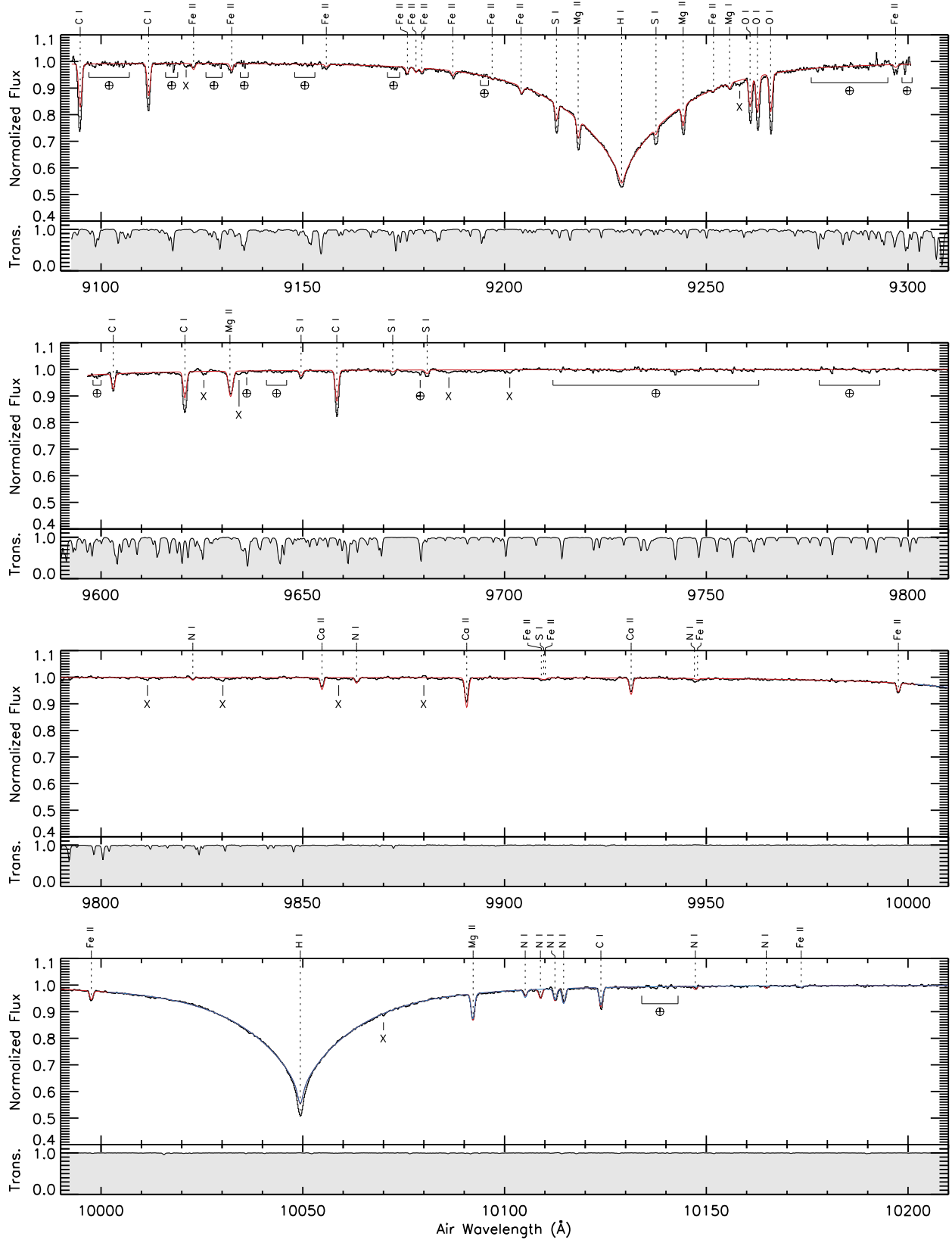}
 \caption{For each wavelength range, two panels present the spectra of
 21 Lyn ({\it upper}) and the atmospheric transmittance ({\it lower}).
 {\it Upper panel:} Continuum-normalized spectrum of 21 Lyn.  The
 synthetic spectra created by ATLAS9 based on the oscillator strengths
 of the VALD database and those of \cite{1999ApJS..124..527M} are
 indicated by red and blue lines, respectively.  The symbole ``X''
 indicates potentially significant features that are not identified in
 the VALD or MB99 in our analysis, while the symbol $\oplus$ indicates
 the spectral regions that show spurious features caused by imperfect
 correction of telluric absorption lines. {\it Lower panel:} Atmospheric
 transmittance derived from the spectrum of the telluric standard star.}
 \label{fig:spectra}
\end{figure*}

\setcounter{figure}{0}
\begin{figure*}[p]
 \epsscale{1.1} \plotone{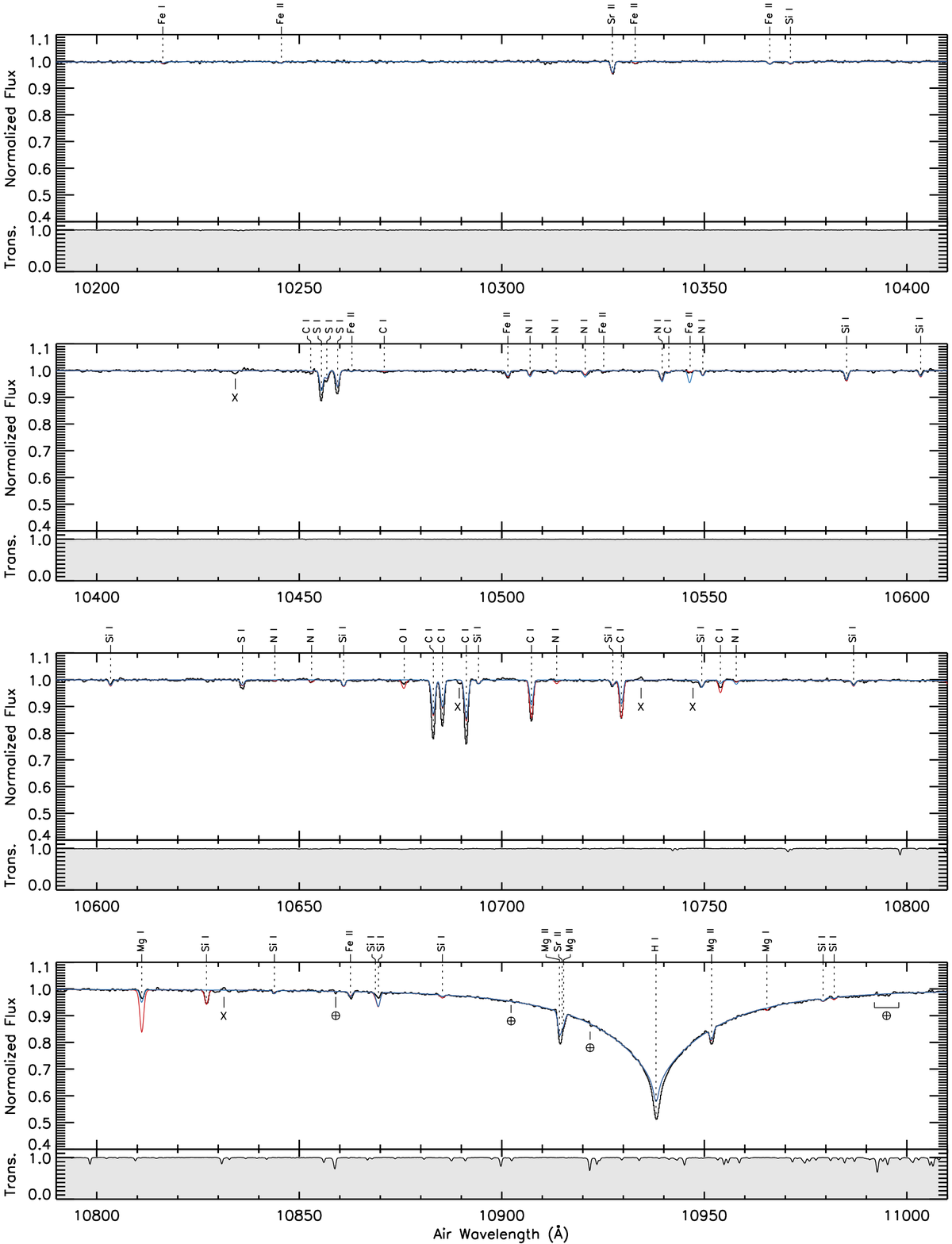}
 \caption{{\it continued.}}
\end{figure*}

\setcounter{figure}{0}
\begin{figure*}[p]
 \epsscale{1.1} \plotone{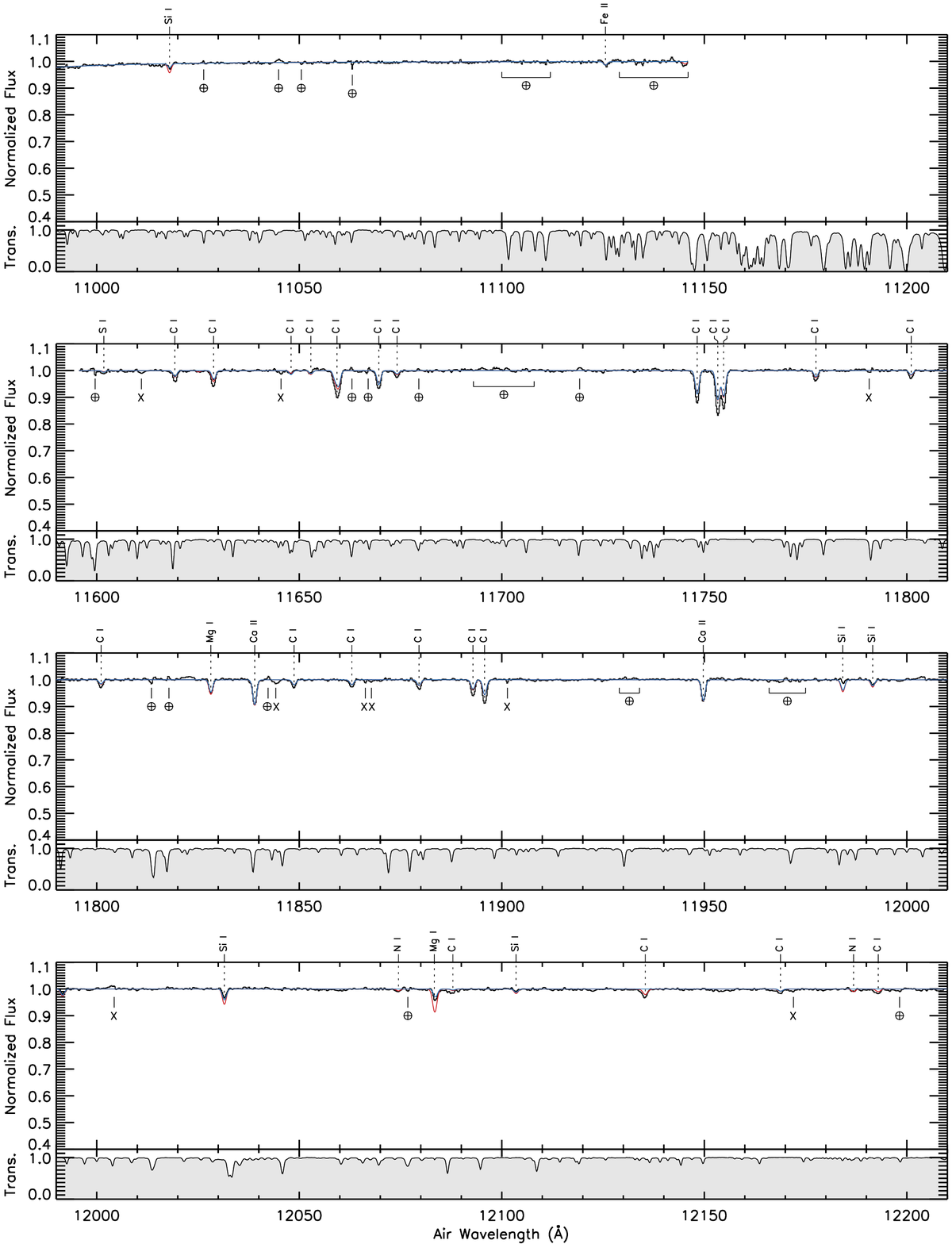}
 \caption{{\it continued.}}
\end{figure*}

\setcounter{figure}{0}
\begin{figure*}[p]
 \epsscale{1.1} \plotone{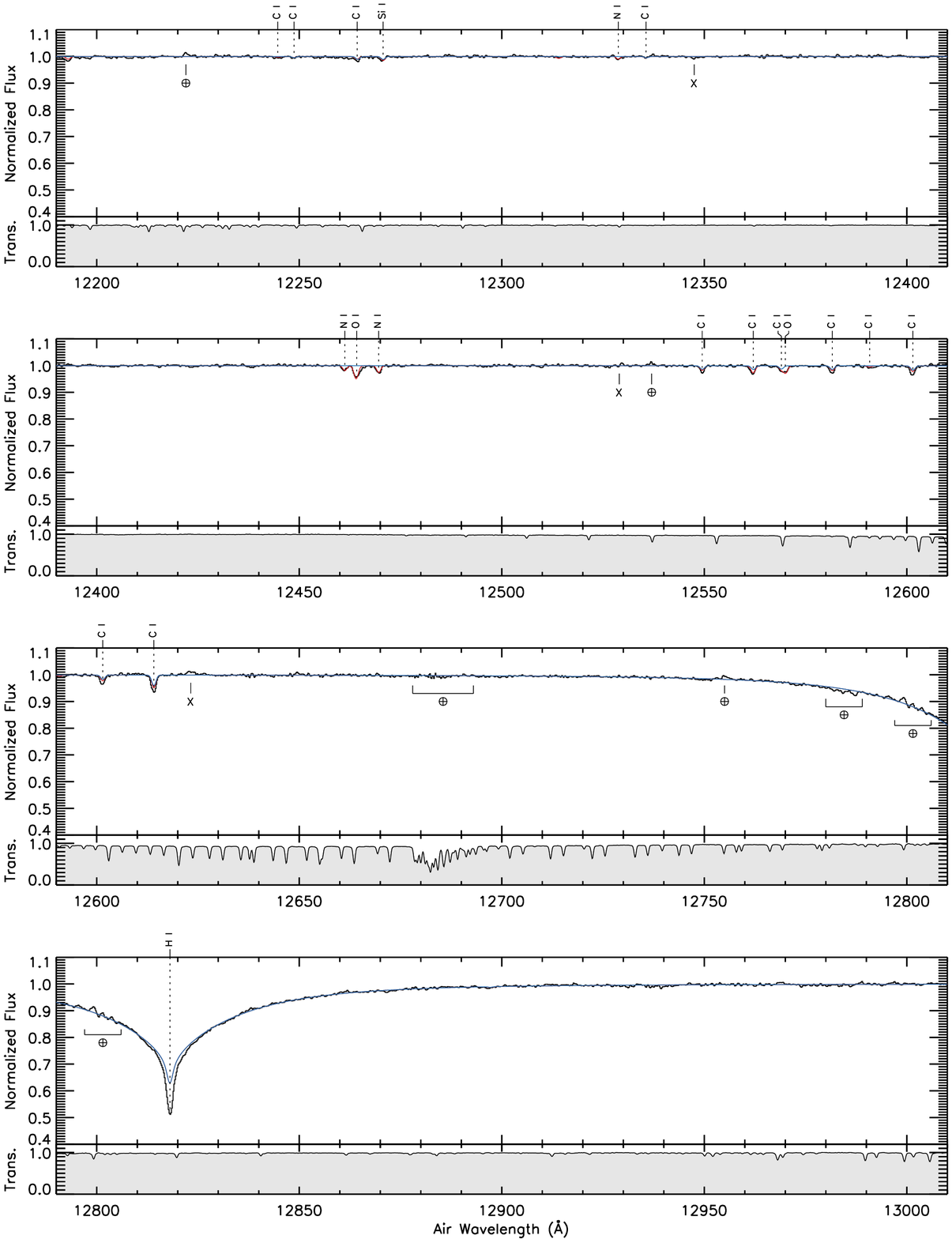}
 \caption{{\it continued.}}
\end{figure*}

\setcounter{figure}{0}
\begin{figure*}[t]
 \epsscale{1.1} \plotone{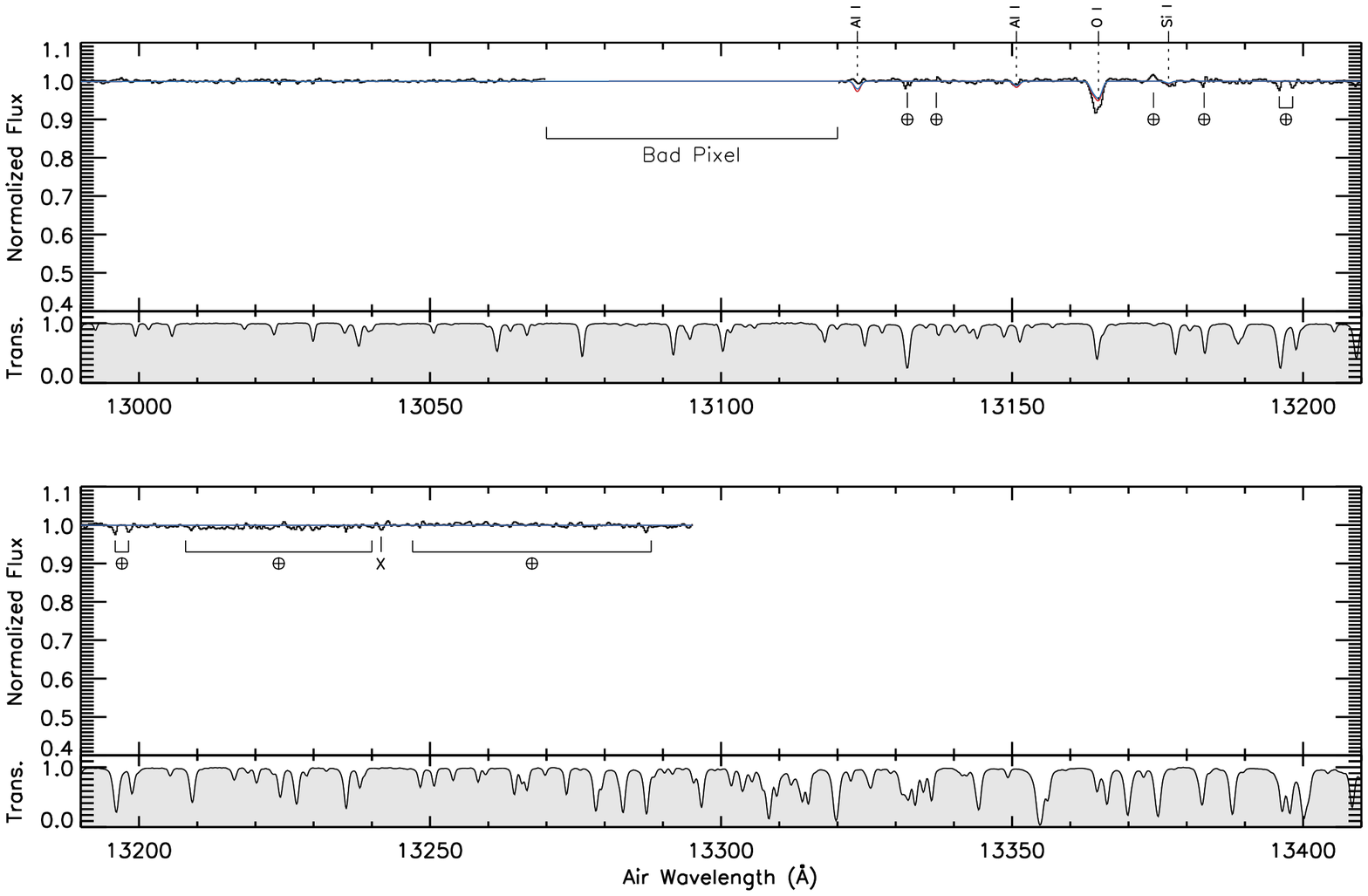}
 \caption{{\it continued.}}
\end{figure*}

\section{Line identification}

\begin{deluxetable}{lr}[t]
 \tablecaption{Parameters of the synthetic spectra for 21 Lyn \label{tab:model_param}}
 \tablehead{
 \colhead{Parameter} & \colhead{Value} \\
 }
 \startdata
 Effective temperature: $T_\mathrm{eff}$\,(K)   & $9520\pm125$ \\
 Surface gravity: $\log g$                      & $3.79\pm0.2$ \\
 Microturbulence: $\xi$\,(km~s$^{-1}$)          & $1.7\pm0.5$  \\
 Rotational velocity: $v \sin i$\,(km~s$^{-1}$) & $18.7\pm0.4$ \\
 \hline
 \multicolumn{2}{c}{Elemental abundance} \\
 \hline
 Carbon: $[\mathrm{C/H}]$ with $\log \varepsilon_{\mathrm{C},\odot}$ = 8.52       & $-0.452\pm0.056$ \\
 Oxygen: $[\mathrm{O/H}]$ with $\log \varepsilon_{\mathrm{O},\odot}$ = 8.83       & $-0.246\pm0.056$ \\
 Magnesium: $[\mathrm{Mg/H}]$ with $\log \varepsilon_{\mathrm{Mg},\odot}$ = 7.58  & $-0.065\pm0.175$ \\
 Silicon: $[\mathrm{Si/H}]$ with $\log \varepsilon_{\mathrm{Si},\odot}$ = 7.55    & $+0.269\pm0.221$ \\
 Calcium: $[\mathrm{Ca/H}]$ with $\log \varepsilon_{\mathrm{Ca},\odot}$ = 6.36    & $-0.057\pm0.417$ \\
 Scandium: $[\mathrm{Sc/H}]$ with $\log \varepsilon_{\mathrm{Sc},\odot}$ = 3.17   & $-0.466\pm0.118$ \\
 Titanium: $[\mathrm{Ti/H}]$ with $\log \varepsilon_{\mathrm{Ti},\odot}$ = 5.02   & $-0.165\pm0.085$ \\
 Chromium: $[\mathrm{Cr/H}]$ with $\log \varepsilon_{\mathrm{Cr},\odot}$ = 5.67   & $+0.055\pm0.078$ \\
 Iron: $[\mathrm{Fe/H}]$ with $\log \varepsilon_{\mathrm{Fe},\odot}$ = 7.50       & $-0.004\pm0.124$ \\
 Nickel: $[\mathrm{Ni/H}]$ with $\log \varepsilon_{\mathrm{Ni},\odot}$ = 6.25     & $+0.287\pm0.136$ \\
 Strontium: $[\mathrm{Sr/H}]$ with $\log \varepsilon_{\mathrm{Sr},\odot}$ = 2.97  & $+0.410\pm0.150$ \\
 Yttrium: $[\mathrm{Y/H}]$ with $\log \varepsilon_{\mathrm{Y},\odot}$ = 2.24      & $+0.313\pm0.105$ \\
 Zirconium: $[\mathrm{Zr/H}]$ with $\log \varepsilon_{\mathrm{Zr},\odot}$ = 2.60  & $+0.550\pm0.097$ \\
 Barium: $[\mathrm{Ba/H}]$ with $\log \varepsilon_{\mathrm{Ba},\odot}$ = 2.13     & $+0.912\pm0.168$ \\
 \enddata
 \tablecomments{The solar abundance for each element, indicated in the
 left column, was adopted from \cite{1998SSRv...85..161G} in both this
 study and \cite{2014AA...562A..84R} who obtained the parameters in
 this table.}
\end{deluxetable}

To identify lines in the observed spectrum, we first created model
spectra of 21 Lyn with ATLAS9 (\citealt{1993sssp.book.....K}).
Fundamental stellar parameters were adopted from
\cite{2014AA...562A..84R}, who derived the following parameters of 21
Lyn from an optical high-resolution spectrum: the effective temperature,
the surface gravity, the microturbulence, $v\sin i$, and the chemical
abundances (C, O, Mg, Si, Ca, Sc, Ti, Cr, Fe, Ni, Sr, Y, Zr, and Ba).
The other elemental abundances were set to the solar values, and the
solar abundance was taken from \cite{1998SSRv...85..161G}.  The adopted
parameters are summarized in Table \ref{tab:model_param}.  The
excitation potentials, transition terms, oscillator strengths ($\log
gf$), and damping constants of the spectral lines used for spectral
synthesis were retrieved from the Vienna Atomic Line Database (VALD;
\citealt{1995A&AS..112..525P,1999A&AS..138..119K,2015PhyS...90e4005R}),
which is a compilation of literature information of lines based mainly
on theoretical calculations.  In addition, the solar-based line catalog
given by \citeauthor{1999ApJS..124..527M}
(\citeyear{1999ApJS..124..527M}; hereafter, MB99) was also used for
spectral synthesis.  Note that their line catalog covers 10,000~\AA\ and
longer; therefore, the synthetic spectrum created with it does not cover
the wavelength range shorter than 1.0~\micron.

\begin{figure}[t]
 \epsscale{1.1} \plotone{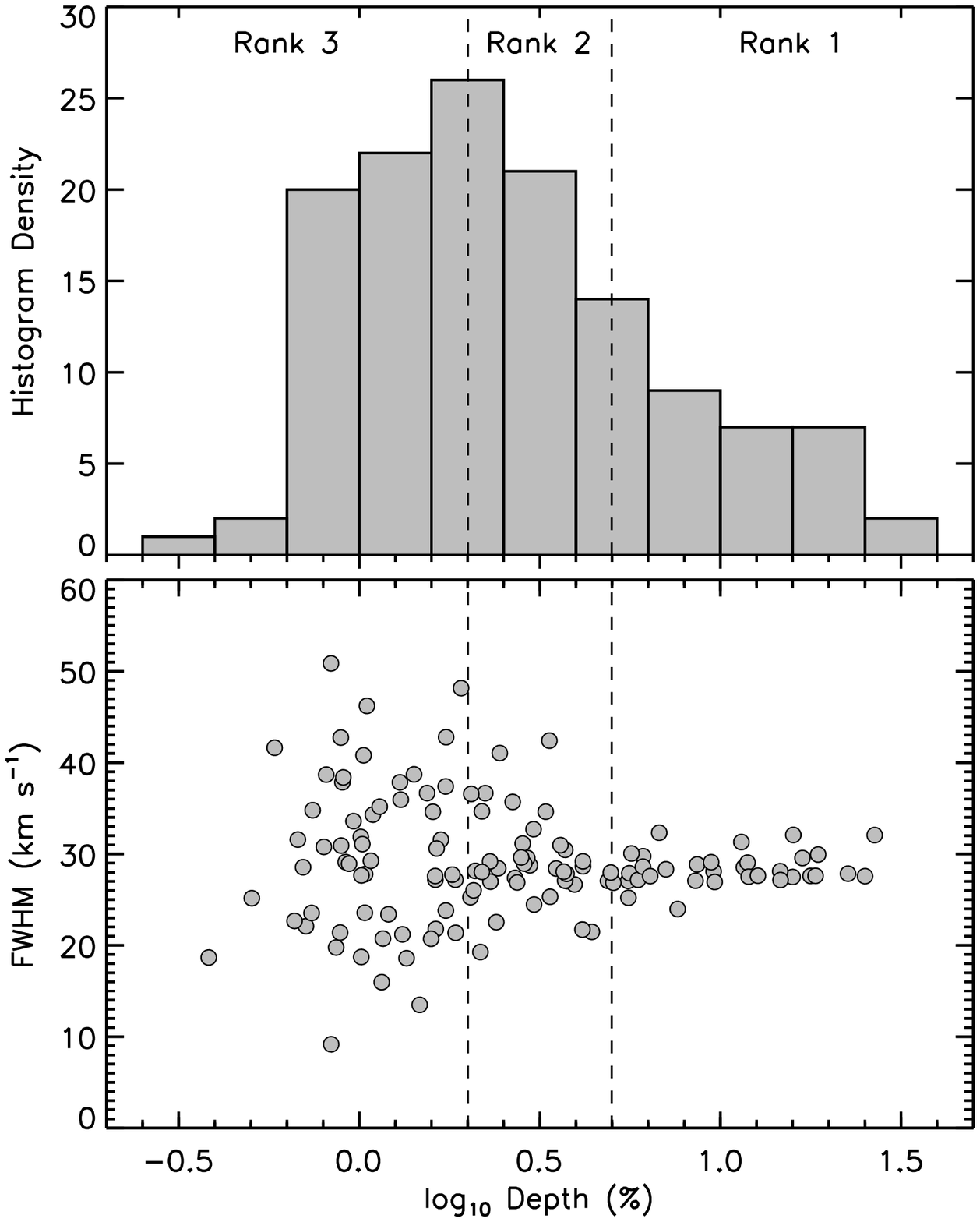}
 \caption{Histogram of the measured line depth ({\it upper}) and
 measured FWHM as a function of the line depth ({\it lower}).  Only the
 unblended lines whose parameters were obtained by fitting single
 Gaussians are included.}  \label{fig:depth_fwhm}
\end{figure}

From visual inspection of the observed and synthetic spectra, we
detected 155 absorption features including very wide hydrogen lines
(Pa\,$\beta$, Pa\,$\gamma$, Pa\,$\delta$, and Pa\,$\zeta$).  Each
absorption feature detected was fitted with a single Gaussian curve to
measure the line depth, the full width at half maximum (FWHM), and the
equivalent width.  The depth was measured in percentage from the
continuum level, 0\% (with no visible absorption) to 100\% (the maximum
depth).  When the absorption feature could not be reproduced with a
single Gaussian curve, we measured the equivalent width by fitting
multiple Gaussian curves and defined the depth of the deepest pixel as
the line depth, while the FWHM was not obtained.  Hydrogen lines were
not measured because their spectral information was destroyed at the
normalization step as described above.  The results of the measurements
for the absorption features are summarized in Table \ref{tab:measure}.

The absorption features were classified according to the strengths in
the following manner.  The upper panel of Figure \ref{fig:depth_fwhm}
shows a histogram of the measured depths for the absorption features
fitted by single Gaussians.  The number of lines increases toward weak
lines, peaks at around 2\%, and then decreases, indicating that the
completeness of line detection changes at a depth around 2\%.  The lower
panel of Figure \ref{fig:depth_fwhm} compares the measured FWHM with the
depth of the same lines.  The scatter in the measured FWHM increases
with decreasing depth, which is simply due to the random noise of the
photon counts.  The scatter is clearly small for the lines stronger than
5\% in depth.  Considering these trends, we classified the detected
features into three strength classes: 1 for lines stronger than 5\% in
depth, 2 for lines with a depth of 2\%--5\%, and 3 for lines weaker than
2\%.  The lines in our line catalog are mainly ranked according to these 
strength classes.

The detected absorption features were then carefully identified by
referring to the synthetic spectra.  When multiple candidates were found
for a single absorption feature, we created synthetic spectra with the
candidate lines ignored one by one.  If the synthetic spectrum did not
change significantly, we concluded that the ignored line does not
contribute to the observed feature and removed it from the candidates.
When more than one line survived after this check made for every
candidate, we judged the absorption feature as blended lines and added
the flag ``B'' to the rank in our line list.  Fine structure lines are
often blended and look like an absorption feature with a single peak;
thus, it is almost impossible to determine which transitions are
actually detected.  In such cases, we included all related transitions
from VALD and MB99 regardless of $\log gf$ and added the flag ``F'' to
the rank of the feature.  Note that we did not try to measure the depth
of each component for those blended or fine structure lines and assigned
the same rank to them.  Some of the absorption features did not have
corresponding lines in the synthetic spectrum based on MB99.  This is
probably due to the difference in the temperatures of 21 Lyn and the
sun.  In those cases, we only used the synthetic spectrum based on VALD
for line identification.

\begin{figure}[!t]
 \epsscale{1.1} \plotone{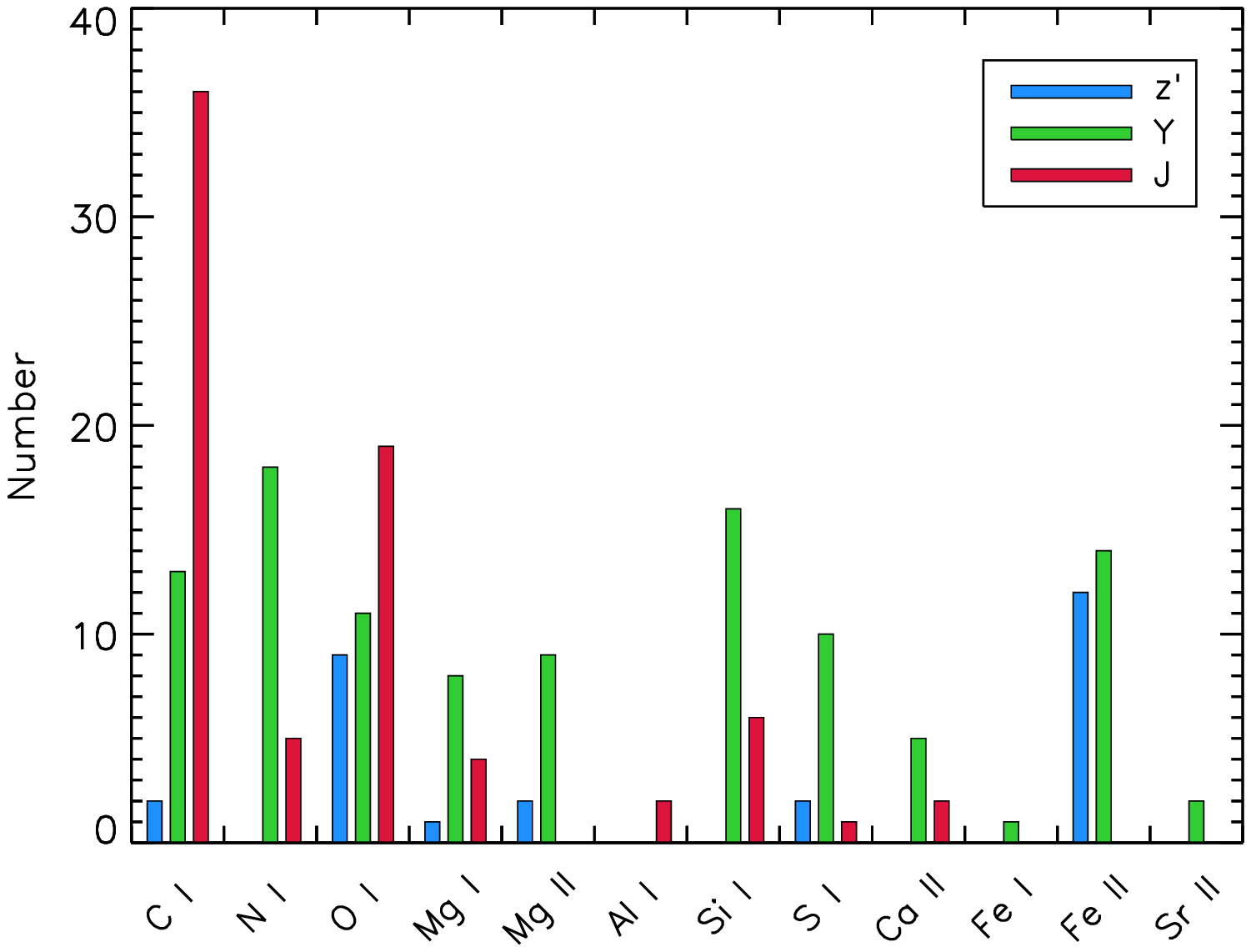}
 \caption{The number of detected lines in the observed spectra of 21
 Lyn in the $z', Y,$ and $J$ bands.}
 \label{fig:detect_line}
\end{figure}

\section{Results}

\begin{figure*}[p]
 \epsscale{1.1} \plotone{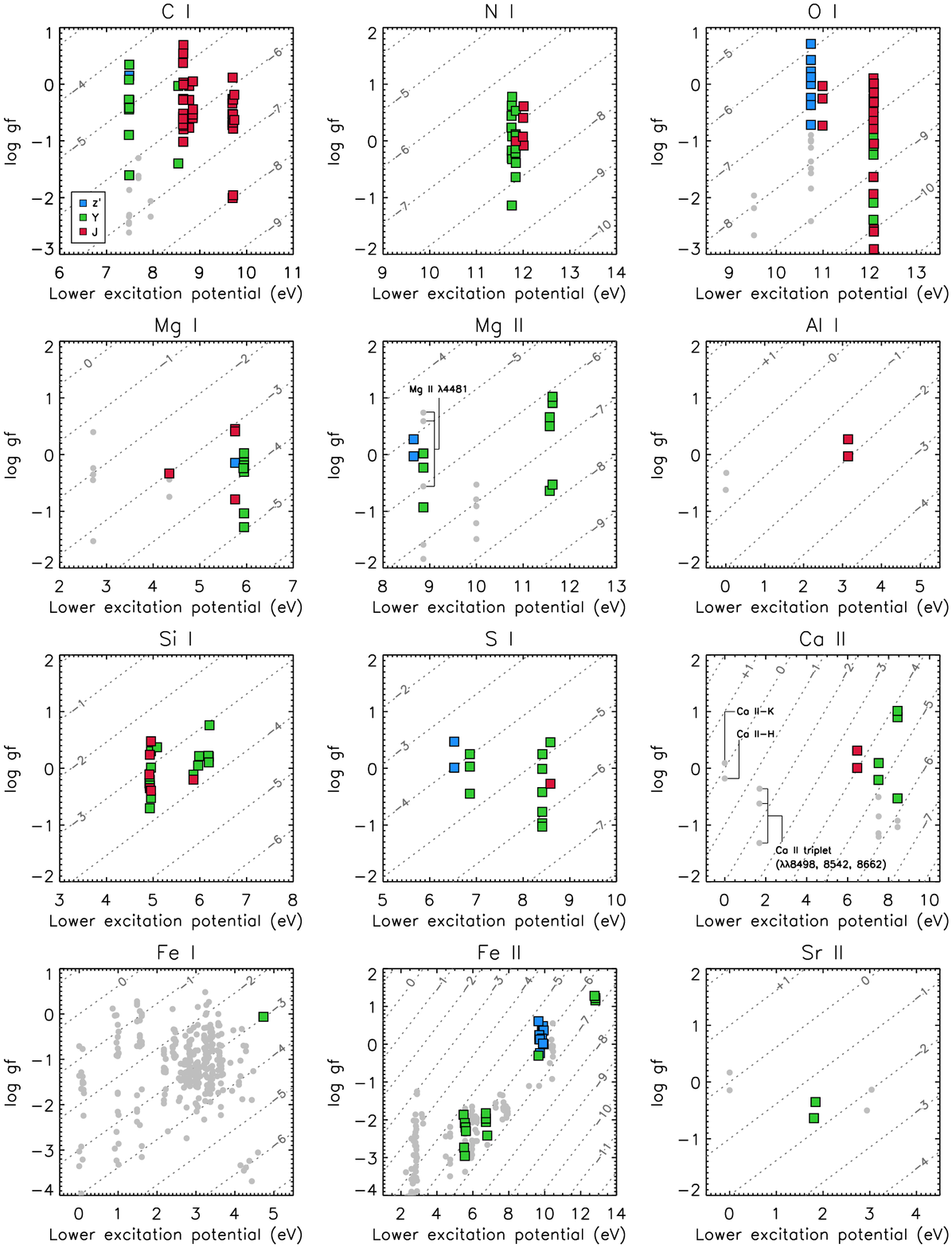} \caption{Distributions of
 the detected lines, indicated by squares (with different colors for
 different bands) in the $\log gf$--EP plane.  The diagonal dotted lines
 show the contours of the line-strength index $X$, and the absorption
 lines are expected to be strong toward the upper-left corner where $X$
 is larger.  The absorption lines in the optical range used for spectral
 classification or abundance analyses in the literature are also plotted
 with gray circles for comparison.}  \label{fig:ep_loggf}
\end{figure*}

As a result of line identification, we conclude that the 155 absorption
features are composed of 219 atomic lines including fine structure lines
with minor contributions.  The wavelength, the element, the $\log gf$
values in VALD and MB99, the excitation potential, the transition term,
and the ranks of the detected lines are summarized in the form of a line
catalog in Table \ref{tab:linelist}.  For reference, the synthetic
spectra based on this line catalog and the photospheric parameters given
in Table \ref{tab:model_param} are compared with the observed spectra in
Figure \ref{fig:spectra}.  The list includes the lines of \ion{H}{1},
\ion{C}{1}, \ion{N}{1}, \ion{O}{1}, \ion{Mg}{1}, \ion{Mg}{2},
\ion{Al}{1}, \ion{Ca}{2}, \ion{Fe}{2}, and \ion{Sr}{2}.  The numbers of
detected lines for individual elements are summarized in Figure
\ref{fig:detect_line}.  Figure \ref{fig:ep_loggf} compares $\log gf$ and
the lower excitation potential of the detected lines for each element.
As was done in \cite{2006ApJ...642..462G}, we introduce a line-strength
index $X=\log gf - \mathrm{EP} \times 5040/(0.86\,T_\mathrm{eff})$,
where EP is the excitation potential of the lower level in
electronvolts, and $T_\mathrm{eff}$ is the effective temperature in
degrees Kelvin.  The diagonal dashed lines in Figure \ref{fig:ep_loggf}
indicate constant $X$ lines, and the absorption lines are expected to be
strong toward the upper-left corner.  For comparison, the optical lines
used for spectral classification or abundance analyses of A-type stars
in the literature (\citealt{1994MNRAS.271..355A,2014AA...562A..84R})
are also plotted.

Besides these identified lines, the spectrum of 21 Lyn show several
features that we could not conclude whether the feature is a stellar
line or just a spurious one for reasons such as weakness, strange line
profiles, and missing from synthetic spectra.  Table \ref{tab:unID}
summarizes these unidentified features with our brief notes.  These
features need to be investigated hopefully with higher-quality spectra
in the future.

\section{Discussion}

  Thanks to the high resolution and high sensitivity of WINERED, we
increased the detected lines of an A-type star in the NIR drastically.
In particular, the lines of various elements in the $Y$ band (see Figure
\ref{fig:detect_line}), a relatively unexplored band, are of scientific
importance for studies of A-type stars.  The richness of the detected
lines suggests the possibility of spectral classification, the
evaluation of chemical peculiarities, and abundance analyses of A-type
stars based solely on the NIR spectra.  Below, we give brief comments on
how the produced line catalog would be useful for various applications.

\subsection{Spectral classification}

In the optical spectrum, the strengths of the \ion{Ca}{2} H \& K lines
relative to Balmer lines have been used for temperature classification
(\citealt{2009ssc..book.....G}).  For the wavelength range of
8500--8750~\AA, which is slightly shorter than our coverage,
\cite{1999A&AS..137..521M} proposed that the combination of \ion{Ca}{2}
triplet lines ($\lambda\lambda8498, 8542, 8662$) and Paschen lines
serves as an indicator of the spectral type between B8 and F8 like
\ion{Ca}{2} H \& K lines.  As an extension of these methods, the
combination of seven \ion{Ca}{2} lines detected in this work and Paschen
lines may also be used for temperature classification in the $Y$ and $J$
bands, which should be investigated in the future using the spectra of
early-type stars with different temperatures.

\subsection{Chemical peculiarities}

\cite{1974ARA&A..12..257P} divided chemically peculiar hot stars into
four groups including Am and Ap stars.  The Am stars are characterized
by weak \ion{Ca}{2} and/or \ion{Sc}{2} lines and enhanced heavy metals.
Although no \ion{Sc}{2} line was detected in 21 Lyn, the presence of the
\ion{Ca}{2} and \ion{Fe}{2} lines detected in this work indicates the
possibility of judging whether or not a target is an Am star solely from
NIR spectra.  On the other hand, the characteristics of the Ap stars are
strong magnetic fields and enhanced abundances of elements such as Si,
Cr, Sr, and Eu.  As can be seen in Figure \ref{fig:detect_line}, many
\ion{Si}{1} lines are present in the NIR.  As for Sr, we detected two
\ion{Sr}{2} lines.  \ion{Sr}{2} $\lambda10914$ is heavily blended with
\ion{Mg}{2} $\lambda\lambda 10914, 10915$ and probably not a good
diagnostic line for chemically peculiar stars.  By contrast, the
moderately strong \ion{Sr}{2} $\lambda10327$, which is fortunately free
from telluric absorption, is not blended with other lines and thus
expected to be useful as a diagnostic line for Ap stars.

\begin{figure}[t]
 \epsscale{1.1} \plotone{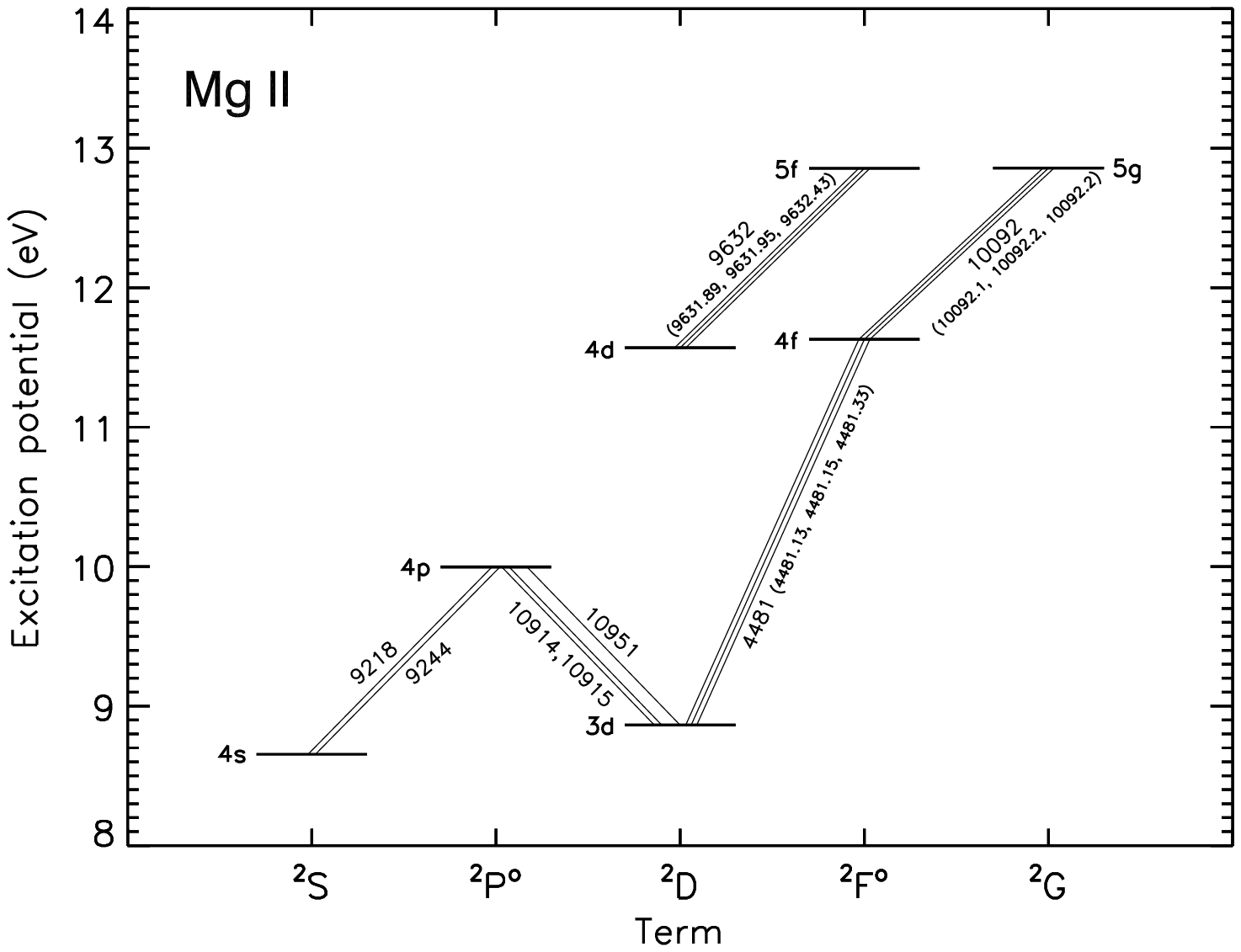}
 \caption{Partial Grotrian diagram of \ion{Mg}{2}.  The transitions of
 the detected lines in this work are illustrated as well as those of
 \ion{Mg}{2} $\lambda4481$ lines, which are known as diagnostic lines
 for $\lambda$ Bootis stars.}
 \label{fig:grotrian_MgII}
\end{figure}

In addition to the Am and Ap stars, $\lambda$ Bootis stars are a group
of A-type stars with chemical peculiarities discovered by
\cite{1943assw.book.....M}.  It is known that the metal absorption lines
of the $\lambda$ Bootis stars are significantly weak compared to other
early-type stars.  In particular, \ion{Mg}{2} $\lambda4481$ lines are
important diagnostic features (\citealt{2009ssc..book.....G}).  Figure
\ref{fig:grotrian_MgII} shows a partial Grotrian diagram of the
\ion{Mg}{2} lines detected in this work and the \ion{Mg}{2}
$\lambda4481$ lines.  As can be seen, \ion{Mg}{2} $\lambda\lambda10914,
10915, 10951$ share the lower term with \ion{Mg}{2} $\lambda4481$ lines
and are thus expected to be similarly useful for diagnostics of
$\lambda$ Bootis stars.  Given that \ion{Mg}{2} $\lambda\lambda10914,
10915$ are heavily blended with \ion{Sr}{2} $\lambda10914$, \ion{Mg}{2}
$\lambda10951$ would be a good proxy for \ion{Mg}{2} $\lambda4481$ lines
in the wavelength range of 0.90--1.35 \micron.

% It is worthwhile to note that the majority of these CP diagnostic lines
% exists at $Y$ band.

\subsection{Abundance analysis}

\begin{figure}[t]
 \epsscale{1.1} \plotone{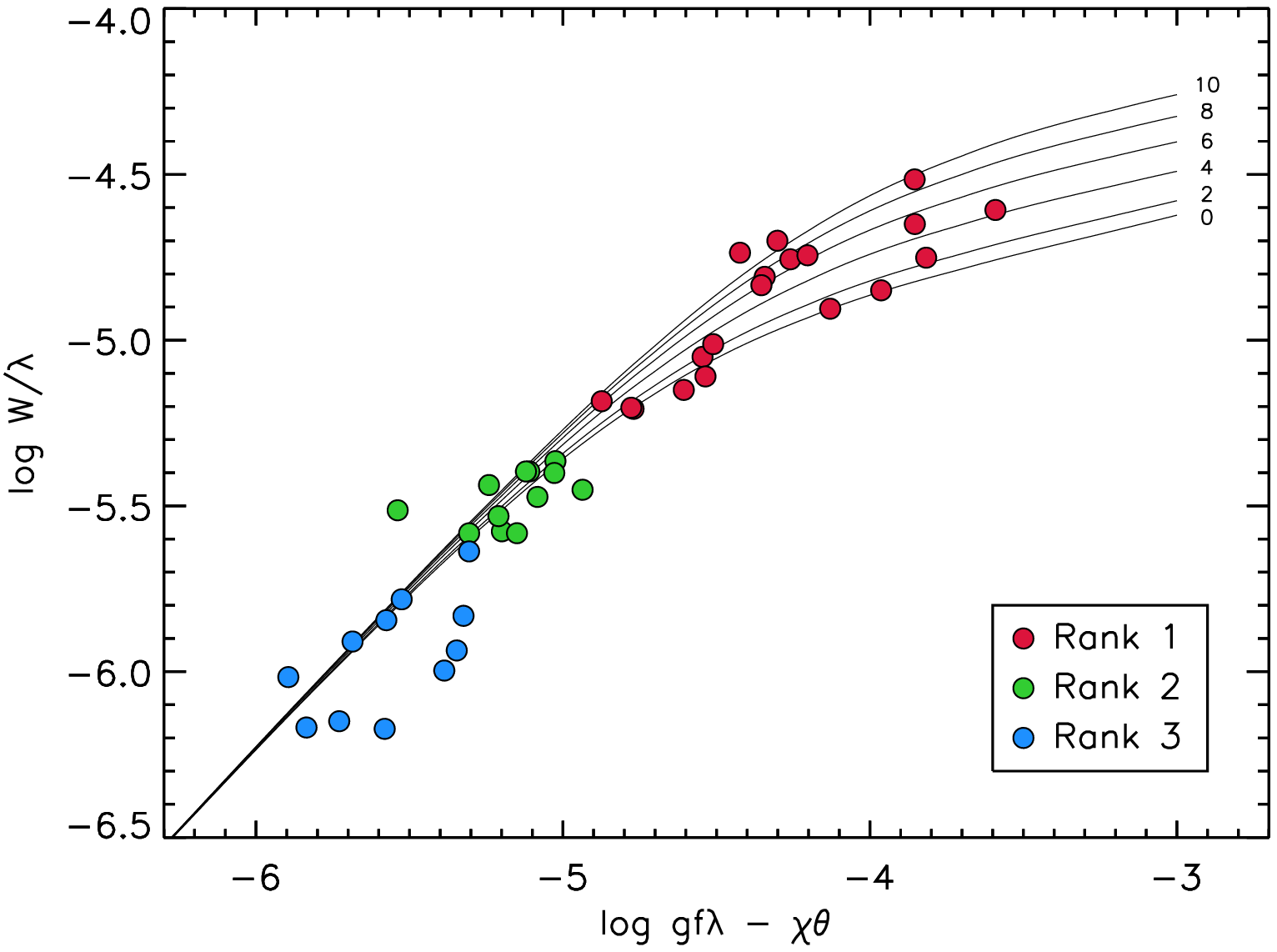}
 \caption{Empirical curve of growth for the \ion{C}{1} lines in 21 Lyn.
 The symbols in the axis titles have their usual meanings (see, e.g.,
 \citealt{1992oasp.book.....G}).  The solid lines indicate the theoretical
 curves of growth created from synthetic spectra with microturbulences
 of $\xi=0,2,4,6,8$, and 10 $\mathrm{km~s^{-1}}$.  Note that the $\log gf$
 values in the VALD database are used here.}  \label{fig:cog}
\end{figure}

A large number of \ion{C}{1}, \ion{N}{1}, \ion{O}{1}, \ion{Mg}{1},
\ion{Mg}{2}, \ion{Si}{1}, \ion{S}{1}, and \ion{Fe}{2} lines in the NIR
spectral range are expected to be useful for measuring CNO abundances
and [$\alpha$/Fe] abundance ratios solely from NIR spectra.  Figure
\ref{fig:cog} shows an empirical curve of growth for the detected
\ion{C}{1} lines in 21 Lyn.  Several strong lines are found in the flat
part of the curve of growth and thus can be used to measure the
microturbulence as well as the chemical abundance.  Having many
absorption lines in the NIR is important for investigating the chemical
abundances of A-type stars with large interstellar extinction.

Although detailed abundance analyses are beyond the scope of this paper,
the large scatter around the curve of growth seen in Figure
\ref{fig:cog} seems to be not only due to measurement errors but also
due to errors in the line database.  Recently,
\cite{2016A&A...585A.143A} reported observationally calibrated $\log gf$
values for NIR \ion{Fe}{1} lines, which were derived from the analysis
of the solar spectrum.  They found that the calibrated $\log gf$ values
were different from the values in the VALD database by more than one dex
for a significant number of NIR \ion{Fe}{1} lines.  These results imply
the importance of observational calibration of $\log gf$ values for
precise abundance analyses in the NIR.  Establishment of line catalogs
for various spectral types, which remains a challenge for our future
research, would be a basis for such calibrations.

% The following lines are used in \cite{1994MNRAS.271..355A} or
% \cite{2014AA...562A..84R} for the abundance analysis of early A-type
% stars in the optical but not detected, at least, on the NIR spectrum of
% 21 Lyn: \ion{C}{2}, \ion{Al}{2}, \ion{Si}{2}, \ion{S}{2}, \ion{Ca}{1},
% \ion{Sc}{2}, \ion{Ti}{1}, \ion{Ti}{2}, \ion{V}{1}, \ion{V}{2},
% \ion{Cr}{1}, \ion{Cr}{2}, \ion{Mn}{1}, \ion{Mn}{2}, \ion{Co}{1},
% \ion{Ni}{1}, \ion{Ni}{2}, \ion{Zn}{1}, \ion{Y}{2}, \ion{Zr}{2},
% \ion{Cd}{1}, \ion{Ba}{2}, \ion{La}{2}, \ion{Ce}{2}, \ion{Pr}{2},
% \ion{Nd}{2}, \ion{Sm}{2}, \ion{Eu}{2}, \ion{Gd}{2}, and \ion{Dy}{2}.

\subsection{A-type stars as telluric standards}

As it is clear in Figure \ref{fig:spectra}, A-type stars are
feature-rich in high-resolution spectroscopy, which should be kept in
mind by observers who plan to use A-type stars as telluric standards.
Our line list would be helpful to distinguish stellar lines from
telluric absorption lines in the observed spectra of A-type stars.
\cite{2018PASP..130g4502S} removed the hydrogen and metal lines from the
WINERED spectra of A0\,V stars to extract the telluric absorption with
the help of our line list and accomplished high-accuracy telluric
correction.

%%%%%%%%%%%%%%%%%%%%%
% Summary
%%%%%%%%%%%%%%%%%%%%%

\section{Summary}

Aiming at extending the coverage of line catalogs in both the wavelength
and spectral type directions, we are carrying out a project to establish
NIR line catalogs based on observations of various types of stars using
our high-dispersion echelle spectrograph, WINERED.  As the first step of
our project, the line catalog of an A-type star was presented in the
current paper.

The spectrum of 21 Lyn, a slowly rotating A0.5\,V star, was obtained by
WINERED in the wavelength range of 0.90--1.35~\micron\ at a resolving
power of $R=28,000$.  After the careful removal of telluric absorption
using a B-type star as a telluric standard, we detected 155 absorption
features including very wide hydrogen lines in the spectrum of 21 Lyn.
Line identification was then performed by visual comparison with
synthetic spectra whose line information was retrieved from the VALD
database and \cite{1999ApJS..124..527M}.  The 155 absorption features
were found to be composed of 219 atomic lines.  Some of these lines may
have only minor contributions but are included as groups of lines that
reproduce $\sim$10 features in our spectrum formed of fine structure
lines.

We compiled the identified 219 atomic lines as a line list, in which
lines are mainly ranked according to the strengths.  Our line list
includes the lines of \ion{H}{1}, \ion{C}{1}, \ion{N}{1}, \ion{O}{1},
\ion{Mg}{1}, \ion{Mg}{2}, \ion{Al}{1}, \ion{Ca}{2}, \ion{Fe}{2}, and
\ion{Sr}{2}.  In particular, the lines of various elements in the $Y$
band, a relatively unexplored band, are of scientific importance for
studies of A-type stars.  The richness of the detected lines suggests
the possibility of spectral classification, the evaluation of chemical
peculiarities, and abundance analyses for A-type stars.  The potential
of such analyses based only on NIR spectra is important for
investigating A-type stars with large interstellar extinction.  Finally,
our line catalog would be helpful to distinguish stellar lines from
telluric absorption lines when A-type stars are used as telluric
standards.

%%%%%%%%%%%%%%%%%%%%%
% acknowledgments
%%%%%%%%%%%%%%%%%%%%%

\acknowledgments

We are grateful to the staff of the Koyama Astronomical Observatory for
their support during our observation.  This study is financially
supported by JSPS KAKENHI (Grant Numbers 16684001, 20340042, 21840052,
and 26287028) and the MEXT Supported Program for the Strategic Research
Foundation at Private Universities, 2008–2012 (No. S0801061) and
2014–2018 (No. S1411028).  N.M., N.K., and H.K. are supported by a JSPS
Grant-in-Aid for Scientific Research (Grant Number 18H01248).  K.F. is
supported by a JSPS Grant-in-Aid for Research Activity Start-up (Grant
Number 16H07323).  S.H. acknowledges support from JSPS through a
Grant-in-Aid for JSPS Fellows (Grant Number 13J10504).  N.K. is
supported by JSPS-DST under the Japan-India Science Cooperative Programs
during 2013--2015 and 2016--2018.  This work has made use of the VALD
database, operated at Uppsala University, the Institute of Astronomy RAS
in Moscow, and the University of Vienna.  We thank the referee,
Prof. Robert Kurucz, for comments that helped us to improve this paper.

\clearpage

\startlongtable
\begin{deluxetable}{lcccccc}
 \tabletypesize{\footnotesize}
 \tablecaption{Measurement of the absorption features in 21 Lyn \label{tab:measure}}
 \tablehead{
 \colhead{ID} &
 \colhead{$\lambda_{\mathrm{air}}$} &
 \colhead{Depth} &
 \colhead{EW} &
 \colhead{FWHM} &
 \colhead{Rank} &
 \colhead{Elements} \\
 & \colhead{(\AA)} & \colhead{(\%)} & \colhead{(m\AA)} & \colhead{(km~s$^{-1}$)}}
 \startdata
  1  & 9094.8 & 26.7 & 277.9 & 32.1 & 1  & \ion{C}{1} \\
  2  & 9111.8 & 18.6 & 181.8 & 29.9 & 1  & \ion{C}{1} \\
  3  & 9122.9 &  1.6 &  14.4 & 27.2 & 3  & \ion{Fe}{2} \\
  4  & 9132.4 &  3.4 &  46.5 & 42.4 & 2  & \ion{Fe}{2} \\
  5  & 9155.8 &  1.9 &  30.2 & 48.2 & 3  & \ion{Fe}{2} \\
  6  & 9175.7 &  2.8 &  30.3 & 31.1 & 2  & \ion{Fe}{2} \\
  7  & 9178.1 &  1.5 &  18.9 & 36.7 & 3  & \ion{Fe}{2} \\
  8  & 9179.5 &  2.3 &  20.8 & 27.0 & 2  & \ion{Fe}{2} \\
  9  & 9187.2 &  2.5 &  34.1 & 41.0 & 2  & \ion{Fe}{2} \\
 10  & 9196.9 &  0.9 &   5.9 & 19.8 & 3  & \ion{Fe}{2} \\
 11  & 9204.1 &  2.7 &  33.1 & 34.5 & 2F & \ion{Fe}{2} \\
 12  & 9212.9 & 11.6 & 128.8 & 28.5 & 1  & \ion{S}{1} \\
 13  & 9218.3 & 11.9 & 144.6 & 29.1 & 1  & \ion{Mg}{2} \\
 14  & 9229.0 & \nodata & \nodata & \nodata & 1 & \ion{H}{1} (Pa\,$\zeta$) \\
 15  & 9237.5 &  6.1 &  80.0 & 29.8 & 1  & \ion{S}{1} \\
 16  & 9244.3 & 11.4 & 138.7 & 31.3 & 1  & \ion{Mg}{2} \\
 17  & 9251.8 &  1.2 &   8.9 & 20.7 & 3  & \ion{Fe}{2} \\
 18  & 9255.8 &  2.2 &  29.3 & 36.7 & 2  & \ion{Mg}{1} \\
 19  & 9260.8 & 17.5 & 168.6 & 27.7 & 1F & \ion{O}{1} \\
 20  & 9262.8 & 20.6 & 211.8 & 29.4 & 1F & \ion{O}{1} \\
 21  & 9266.0 & 23.0 & 224.2 & 28.1 & 1F & \ion{O}{1} \\
 22  & 9296.9 &  5.7 &  56.8 & 30.1 & 1  & \ion{Fe}{2} \\
 23  & 9603.0 &  7.6 &  62.9 & 24.0 & 1  & \ion{C}{1} \\
 24  & 9620.8 & 15.9 & 176.6 & 32.1 & 1  & \ion{C}{1} \\
 25  & 9631.9 &  8.3 & 109.2 & 37.9 & 1F & \ion{Mg}{2} \\
 26  & 9649.6 &  3.0 &  29.4 & 28.8 & 2  & \ion{S}{1} \\
 27  & 9658.4 & 17.8 & 169.4 & 27.6 & 1  & \ion{C}{1} \\
 28  & 9672.3 &  1.7 &  21.7 & \nodata & 3F & \ion{S}{1} \\
 29  & 9680.8 &  2.7 &  30.0 & 32.7 & 2F & \ion{S}{1} \\
 30  & 9822.8 &  1.1 &  13.0 & 34.3 & 3  & \ion{N}{1} \\
 31  & 9854.8 &  3.8 &  35.5 & 27.0 & 2  & \ion{Ca}{2} \\
 32  & 9863.3 &  1.7 &  18.7 & 31.6 & 3  & \ion{N}{1} \\
 33  & 9890.6 &  8.8 &  86.4 & 28.0 & 1F & \ion{Ca}{2} \\
 34  & 9909.7 &  1.3 &  29.2 & \nodata & 3B & \ion{Fe}{2}, \ion{S}{1} \\
 35  & 9931.4 &  5.6 &  53.0 & 27.1 & 1 & \ion{Ca}{2} \\
 36  & 9947.1 &  0.8 &  10.0 & \nodata & 3B & \ion{N}{1}, \ion{Fe}{2} \\
 37  & 9997.6 &  3.9 &  38.2 & 26.7 & 2 & \ion{Fe}{2} \\
 38  & 10049.4 & \nodata & \nodata & \nodata & 1 & \ion{H}{1} (Pa\,$\delta$) \\
 39  & 10092.1 & 10.0 & 115.0 & 31.1 & 1F & \ion{Mg}{2} \\
 40  & 10105.1 &  2.4 &  19.8 & 22.5 & 2  & \ion{N}{1} \\
 41  & 10108.9 &  3.4 &  31.2 & 25.3 & 2  & \ion{N}{1} \\
 42  & 10112.5 &  4.9 &  48.2 & 27.1 & 2  & \ion{N}{1} \\
 43  & 10114.6 &  5.1 &  49.4 & 26.9 & 1  & \ion{N}{1} \\
 44  & 10123.9 &  8.6 &  90.2 & 28.9 & 1  & \ion{C}{1} \\
 45  & 10147.3 &  1.4 &   9.1 & 18.6 & 3  & \ion{N}{1} \\
 46  & 10164.8 &  0.9 &   9.7 & 29.1 & 3  & \ion{N}{1} \\
 47  & 10173.5 &  0.8 &  15.4 & 50.9 & 3  & \ion{Fe}{2} \\
 48  & 10216.3 &  1.0 &  11.8 & 33.6 & 3  & \ion{Fe}{1} \\
 49  & 10245.6 &  0.6 &   8.8 & 41.6 & 3  & \ion{Fe}{2} \\
 50  & 10327.3 &  5.0 &  50.9 & 28.0 & 2  & \ion{Sr}{2} \\
 51  & 10332.9 &  0.9 &  12.5 & 37.8 & 3  & \ion{Fe}{2} \\
 52  & 10366.2 &  1.0 &  11.9 & 31.9 & 3  & \ion{Fe}{2} \\
 53  & 10371.3 &  0.8 &   9.1 & 30.8 & 3  & \ion{Si}{1} \\
 54  & 10452.8 &  1.0 &   7.1 & 18.7 & 3  & \ion{C}{1} \\
 55  & 10455.5 & 12.0 & 122.9 & 27.5 & 1  & \ion{S}{1} \\
 56  & 10456.8 &  4.2 &  44.4 & 28.7 & 2  & \ion{S}{1} \\
 57  & 10459.5 &  9.6 & 100.0 & 28.1 & 1  & \ion{S}{1} \\
 58  & 10463.0 &  0.7 &   5.9 & 22.1 & 3  & \ion{Fe}{2} \\
 59  & 10471.0 &  0.7 &   8.0 & 31.6 & 3  & \ion{C}{1} \\
 60  & 10501.5 &  2.2 &  28.3 & 34.7 & 2  & \ion{Fe}{2} \\
 61  & 10507.0 &  1.6 &  13.3 & 21.8 & 3  & \ion{N}{1} \\
 62  & 10513.4 &  1.0 &  10.8 & 27.8 & 3  & \ion{N}{1} \\
 63  & 10520.6 &  1.7 &  27.8 & 42.8 & 3  & \ion{N}{1} \\
 64  & 10525.2 &  0.9 &  10.3 & 30.9 & 3  & \ion{Fe}{2} \\
 65  & 10539.6 &  3.7 &  37.7 & 27.1 & 2  & \ion{N}{1} \\
 66  & 10541.2 &  0.9 &  10.2 & 28.9 & 3  & \ion{C}{1} \\
 67  & 10546.3 &  1.0 &  15.7 & 40.8 & 3  & \ion{Fe}{2} \\
 68  & 10549.6 &  1.8 &  14.8 & 21.4 & 3  & \ion{N}{1} \\
 69  & 10585.1 &  3.8 &  39.2 & 27.8 & 2  & \ion{Si}{1} \\
 70  & 10603.4 &  2.0 &  19.3 & 25.3 & 2  & \ion{Si}{1} \\
 71  & 10636.0 &  3.5 &  37.7 & 28.4 & 2  & \ion{S}{1} \\
 72  & 10644.0 &  0.5 &   4.8 & 25.2 & 3  & \ion{N}{1} \\
 73  & 10653.0 &  1.4 &  20.6 & 38.7 & 3  & \ion{N}{1} \\
 74  & 10661.0 &  2.4 &  26.1 & 28.4 & 2  & \ion{Si}{1} \\
 75  & 10675.7 &  2.0 &  34.4 & 46.5 & 3F & \ion{O}{1} \\
 76  & 10683.1 & 22.6 & 239.4 & 27.8 & 1  & \ion{C}{1} \\
 77  & 10685.3 & 18.3 & 192.7 & 27.6 & 1  & \ion{C}{1} \\
 78  & 10691.2 & 25.1 & 264.1 & 27.6 & 1  & \ion{C}{1} \\
 79  & 10694.3 &  2.1 &  22.4 & 28.2 & 2  & \ion{Si}{1} \\
 80  & 10707.3 & 15.8 & 166.3 & 27.5 & 1  & \ion{C}{1} \\
 81  & 10713.5 &  0.8 &  11.9 & 38.7 & 3  & \ion{N}{1} \\
 82  & 10727.4 &  2.3 &  25.7 & 29.2 & 2  & \ion{Si}{1} \\
 83  & 10729.5 & 14.6 & 157.2 & 28.1 & 1  & \ion{C}{1} \\
 84  & 10749.4 &  2.1 &  20.8 & 26.0 & 2  & \ion{Si}{1} \\
 85  & 10754.0 &  2.9 &  33.0 & 29.6 & 2  & \ion{C}{1} \\
 86  & 10757.9 &  0.4 &   2.7 & 18.7 & 3  & \ion{N}{1} \\
 87  & 10786.8 &  1.6 &  17.2 & 27.6 & 3  & \ion{Si}{1} \\
 88  & 10811.1 &  3.7 &  48.7 & 34.0 & 2F & \ion{Mg}{1} \\
 89  & 10827.1 &  5.6 &  60.1 & 27.9 & 1  & \ion{Si}{1} \\
 90  & 10843.9 &  1.2 &   7.1 & 16.0 & 3  & \ion{Si}{1} \\
 91  & 10862.7 &  2.7 &  36.9 & 35.7 & 2  & \ion{Fe}{2} \\
 92  & 10868.8 &  0.8 &   3.0 &  9.2 & 3  & \ion{Si}{1} \\
 93  & 10869.5 &  2.2 &  16.3 & 19.3 & 2  & \ion{Si}{1} \\
 94  & 10885.3 &  0.9 &  15.0 & 42.7 & 3  & \ion{Si}{1} \\
 95  & 10914.2 & 11.9 & 192.3 & \nodata & 1B & \ion{Mg}{2}, \ion{Sr}{2} \\
 96  & 10938.1 & \nodata & \nodata & \nodata & 1 & \ion{H}{1} (Pa\,$\gamma$) \\
 97  & 10951.8 &  5.9 &  73.7 & 27.2 & 1  & \ion{Mg}{2} \\
 98  & 10965.5 &  1.0 &  11.7 & 27.7 & 3  & \ion{Mg}{1} \\
 99  & 10979.3 &  1.0 &  12.8 & 31.1 & 3  & \ion{Si}{1} \\
 100 & 10982.1 &  0.7 &   7.0 & 23.5 & 3  & \ion{Si}{1} \\
 101 & 11018.0 &  2.2 &  24.1 & 28.0 & 2  & \ion{Si}{1} \\
 102 & 11125.6 &  1.6 &  21.9 & 34.6 & 3  & \ion{Fe}{2} \\
 103 & 11601.8 &  1.3 &  22.0 & 37.8 & 3  & \ion{S}{1} \\
 104 & 11619.3 &  4.4 &  39.1 & 21.5 & 2  & \ion{C}{1} \\
 105 & 11628.8 &  6.1 &  72.2 & 28.6 & 1  & \ion{C}{1} \\
 106 & 11648.0 &  1.3 &  19.3 & 35.9 & 3  & \ion{C}{1} \\
 107 & 11652.8 &  1.6 &  13.5 & 20.7 & 3  & \ion{C}{1} \\
 108 & 11659.7 & 10.4 & 171.8 & 40.2 & 1F & \ion{C}{1} \\
 109 & 11669.6 &  7.1 &  82.6 & 28.3 & 1  & \ion{C}{1} \\
 110 & 11674.1 &  2.7 &  30.5 & 27.4 & 2  & \ion{C}{1} \\
 111 & 11748.2 & 12.7 & 146.2 & 27.6 & 1  & \ion{C}{1} \\
 112 & 11753.3 & 16.9 & 208.5 & 29.6 & 1  & \ion{C}{1} \\
 113 & 11754.8 & 14.7 & 166.2 & 27.2 & 1  & \ion{C}{1} \\
 114 & 11777.5 &  4.2 &  50.8 & 29.2 & 2  & \ion{C}{1} \\
 115 & 11801.1 &  3.7 &  43.1 & 28.1 & 2  & \ion{C}{1} \\
 116 & 11828.2 &  5.6 &  58.6 & 25.2 & 1  & \ion{Mg}{1} \\
 117 & 11839.0 &  9.6 & 109.4 & 26.9 & 1  & \ion{Ca}{2} \\
 118 & 11848.7 &  3.0 &  31.4 & 24.5 & 2  & \ion{C}{1} \\
 119 & 11863.0 &  2.9 &  34.9 & 28.9 & 2  & \ion{C}{1} \\
 120 & 11879.6 &  3.7 &  47.7 & 30.5 & 2  & \ion{C}{1} \\
 121 & 11892.9 &  6.4 &  74.4 & 27.6 & 1  & \ion{C}{1} \\
 122 & 11895.8 &  9.4 & 115.7 & 29.1 & 1  & \ion{C}{1} \\
 123 & 11949.7 &  8.5 &  97.9 & 27.1 & 1  & \ion{Ca}{2} \\
 124 & 11984.2 &  1.5 &   8.4 & 13.5 & 3  & \ion{Si}{1} \\
 125 & 11991.6 &  1.8 &  21.4 & 27.2 & 3  & \ion{Si}{1} \\
 126 & 12031.5 &  4.1 &  38.3 & 21.7 & 2  & \ion{Si}{1} \\
 127 & 12074.5 &  1.1 &  20.8 & 46.2 & 3  & \ion{N}{1} \\
 128 & 12083.6 &  4.4 &  54.4 & 28.9 & 2F & \ion{Mg}{1} \\
 129 & 12087.9 &  1.1 &  17.3 & 35.2 & 3  & \ion{C}{1} \\
 130 & 12103.5 &  1.0 &  10.5 & 23.6 & 3  & \ion{Si}{1} \\
 131 & 12135.4 &  3.0 &  42.9 & 32.7 & 2  & \ion{C}{1} \\
 132 & 12168.8 &  1.2 &  12.3 & 23.4 & 3  & \ion{C}{1} \\
 133 & 12186.8 &  0.7 &   6.5 & 22.7 & 3  & \ion{N}{1} \\
 134 & 12192.9 &  1.7 &  28.1 & 37.4 & 3  & \ion{C}{1} \\
 135 & 12244.9 &  0.6 &  14.4 & 55.9 & 3B & \ion{C}{1} \\
 136 & 12248.7 &  0.7 &   8.7 & 28.6 & 3  & \ion{C}{1} \\
 137 & 12264.3 &  1.7 &  18.1 & 23.8 & 3  & \ion{C}{1} \\
 138 & 12270.7 &  1.6 &  21.9 & 30.6 & 3  & \ion{Si}{1} \\
 139 & 12328.8 &  1.1 &  13.8 & 29.3 & 3  & \ion{N}{1} \\
 140 & 12335.6 &  0.9 &   8.3 & 21.4 & 3  & \ion{C}{1} \\
 141 & 12461.3 &  1.8 &  22.3 & 27.7 & 3  & \ion{N}{1} \\
 142 & 12463.8 &  4.3 &  77.9 & 41.3 & 2F & \ion{O}{1} \\
 143 & 12469.6 &  2.8 &  36.8 & 29.6 & 2  & \ion{N}{1} \\
 144 & 12549.5 &  2.7 &  32.8 & 26.9 & 2  & \ion{C}{1} \\
 145 & 12562.1 &  3.5 &  47.1 & 29.9 & 2B & \ion{C}{1} \\
 146 & 12569.0 &  2.5 &  62.9 & \nodata & 2B & \ion{C}{1}, \ion{O}{1} \\
 147 & 12581.6 &  3.3 &  50.5 & 34.6 & 2  & \ion{C}{1} \\
 148 & 12590.8 &  0.9 &  15.5 & 38.4 & 3  & \ion{C}{1} \\
 149 & 12601.5 &  3.6 &  50.1 & 31.0 & 2  & \ion{C}{1} \\
 150 & 12614.1 &  6.8 &  98.1 & 32.3 & 1  & \ion{C}{1} \\
 151 & 12818.1 & \nodata & \nodata & \nodata & 1 & \ion{H}{1} (Pa\,$\beta$) \\
 152 & 13123.4 &  1.3 &  13.0 & 21.2 & 3  & \ion{Al}{1} \\
 153 & 13150.6 &  0.7 &  12.1 & 34.8 & 3  & \ion{Al}{1} \\
 154 & 13163.9 &  9.0 & 179.9 & 42.8 & 1F & \ion{O}{1} \\
 155 & 13176.9 &  2.0 &  34.8 & 36.6 & 2  & \ion{Si}{1} \\
 \enddata
\end{deluxetable}

\clearpage
\startlongtable
\begin{deluxetable*}{cccccccr@{ -- }lcc}
 \tabletypesize{\footnotesize}
 \tablecaption{Line catalog \label{tab:linelist}}
 \tablehead{
 \multicolumn{2}{c}{Wavelength (\AA)} &
 \colhead{Element} &
 \multicolumn{2}{c}{$\log\,gf$} &
 \multicolumn{2}{c}{EP (eV)} &
 \multicolumn{2}{c}{Term} &
 \colhead{Rank} &
 \colhead{Notes} \\
 \cmidrule(r){1-2} \cmidrule(r){4-5} \cmidrule(r){6-7} \cmidrule(r){8-9}
 \colhead{Air} &
 \colhead{Vacuum} &
 & \colhead{VALD} & \colhead{MB99} & \colhead{lower} &
 \colhead{upper} &
 \colhead{lower} & \colhead{upper} & }
 \startdata
 9094.8287 & 9097.3252 & \ion{C}{1}  &  $+0.151$ &   \nodata &  7.4878 &  8.8507 & $^3P^o_{2}$          & $^3P_{2}$            & 1  & ID=1 \\
 9111.7986 & 9114.2997 & \ion{C}{1}  &  $-0.297$ &   \nodata &  7.4878 &  8.8481 & $^3P^o_{2}$          & $^3P_{1}$            & 1  & ID=2 \\
 9122.9348 & 9125.4389 & \ion{Fe}{2} &  $+0.357$ &   \nodata &  9.8492 & 11.2079 & e\,$^4D_{7/2}$       & $^4D^o_{7/2}$        & 3  & ID=3 \\
 9132.3853 & 9134.8919 & \ion{Fe}{2} &  $+0.426$ &   \nodata &  9.8492 & 11.2065 & e\,$^4D_{7/2}$       & $^4F^o_{9/2}$        & 2  & ID=4 \\
 9155.8239 & 9158.3369 & \ion{Fe}{2} &  $-0.236$ &   \nodata &  9.7359 & 11.0897 & e\,$^6D_{5/2}$       & $^6P^o_{5/2}$        & 3  & ID=5 \\
 9175.9196 & 9178.4380 & \ion{Fe}{2} &  $+0.479$ &   \nodata &  9.9045 & 11.2554 & e\,$^4D_{5/2}$       & $^4F^o_{7/2}$        & 2  & ID=6 \\
 9178.0584 & 9180.5775 & \ion{Fe}{2} &  $+0.362$ &   \nodata &  9.9408 & 11.2913 & e\,$^4D_{3/2}$       & $^4F^o_{5/2}$        & 3  & ID=7 \\
 9179.4919 & 9182.0113 & \ion{Fe}{2} &  $+0.128$ &   \nodata &  9.7002 & 11.0505 & e\,$^6D_{7/2}$       & $^6P^o_{7/2}$        & 2  & ID=8 \\
 9187.1828 & 9189.7043 & \ion{Fe}{2} &  $+0.242$ &   \nodata &  9.7002 & 11.0494 & e\,$^6D_{7/2}$       & $^6D^o_{5/2}$        & 2  & ID=9 \\
 9196.9217 & 9199.4458 & \ion{Fe}{2} &  $-0.002$ &   \nodata &  9.9408 & 11.2885 & e\,$^4D_{3/2}$       & $^4D^o_{3/2}$        & 3  & ID=10 \\
 9204.0952 & 9206.6213 & \ion{Fe}{2} &  $+0.608$ &   \nodata &  9.6536 & 11.0003 & e\,$^6D_{9/2}$       & $^6D^o_{9/2}$        & 2F & ID=11 \\
 9204.6172 & 9207.1434 & \ion{Fe}{2} &  $+0.151$ &   \nodata &  9.8492 & 11.1959 & e\,$^4D_{7/2}$       & $^6F^o_{7/2}$        & 2F & ID=11 \\
 9212.8630 & 9215.3914 & \ion{S}{1}  &  $+0.470$ &   \nodata &  6.5245 &  7.8699 & $^5S^o_{2}$          & $^5P_{3}$            & 1  & ID=12 \\
 9218.2500 & 9220.7799 & \ion{Mg}{2} &  $+0.270$ &   \nodata &  8.6547 &  9.9993 & $^2S_{1/2}$          & $^2P^o_{3/2}$        & 1  & ID=13 \\
 9229.0170 & 9231.5498 & \ion{H}{1} (Pa\,$\zeta$)  &  $-0.735$ & \nodata & 12.0875 & 13.4306 & $n=3$                & $n=9$ & 1 & ID=14 \\
 9237.5380 & 9240.0731 & \ion{S}{1}  &  $+0.010$ &   \nodata &  6.5245 &  7.8663 & $^5S^o_{2}$          & $^5P_{1}$            & 1  & ID=15 \\
 9244.2650 & 9246.8019 & \ion{Mg}{2} &  $-0.030$ &   \nodata &  8.6547 &  9.9955 & $^2S_{1/2}$          & $^2P^o_{1/2}$        & 1  & ID=16 \\
 9251.7872 & 9254.3262 & \ion{Fe}{2} &  $+0.125$ &   \nodata &  9.7359 & 11.0757 & e\,$^6D_{5/2}$       & $^6D^o_{3/2}$        & 3  & ID=17 \\
 9255.7780 & 9258.3181 & \ion{Mg}{1} &  $-0.146$ &   \nodata &  5.7532 &  7.0924 & $^1D_{2}$            & $^1F^o_{3}$          & 2  & ID=18 \\
 9260.8060 & 9263.3474 & \ion{O}{1}  &  $-0.241$ &   \nodata & 10.7402 & 12.0787 & $^5P_{1}$            & $^5D^o_{0}$          & 1F & ID=19 \\
 9260.8480 & 9263.3894 & \ion{O}{1}  &  $+0.110$ &   \nodata & 10.7402 & 12.0787 & $^5P_{1}$            & $^5D^o_{1}$          & 1F & ID=19 \\
 9260.9360 & 9263.4775 & \ion{O}{1}  &  $+0.002$ &   \nodata & 10.7402 & 12.0786 & $^5P_{1}$            & $^5D^o_{2}$          & 1F & ID=19 \\
 9262.5820 & 9265.1239 & \ion{O}{1}  &  $-0.368$ &   \nodata & 10.7405 & 12.0787 & $^5P_{2}$            & $^5D^o_{1}$          & 1F & ID=20 \\
 9262.6700 & 9265.2119 & \ion{O}{1}  &  $+0.224$ &   \nodata & 10.7405 & 12.0786 & $^5P_{2}$            & $^5D^o_{2}$          & 1F & ID=20 \\
 9262.7760 & 9265.3180 & \ion{O}{1}  &  $+0.427$ &   \nodata & 10.7405 & 12.0786 & $^5P_{2}$            & $^5D^o_{3}$          & 1F & ID=20 \\
 9265.8260 & 9268.3688 & \ion{O}{1}  &  $-0.718$ &   \nodata & 10.7409 & 12.0786 & $^5P_{3}$            & $^5D^o_{2}$          & 1F & ID=21 \\
 9265.9320 & 9268.4748 & \ion{O}{1}  &  $+0.125$ &   \nodata & 10.7409 & 12.0786 & $^5P_{3}$            & $^5D^o_{3}$          & 1F & ID=21 \\
 9266.0060 & 9268.5488 & \ion{O}{1}  &  $+0.712$ &   \nodata & 10.7409 & 12.0786 & $^5P_{3}$            & $^5D^o_{4}$          & 1F & ID=21 \\
 9296.9197 & 9299.4709 & \ion{Fe}{2} &  $+0.018$ &   \nodata & 9.9045  & 11.2378 & e\,$^4D_{5/2}$       & $^4D^o_{5/2}$        & 1  & ID=22 \\
 9603.0294 & 9605.6635 & \ion{C}{1}  &  $-0.896$ &   \nodata &  7.4804 &  8.7711 & $^3P^o_{0}$          & $^3S_{1}$            & 1  & ID=23 \\
 9620.7822 & 9623.4211 & \ion{C}{1}  &  $-0.445$ &   \nodata &  7.4828 &  8.7711 & $^3P^o_{1}$          & $^3S_{1}$            & 1  & ID=24 \\
 9631.8910 & 9634.5329 & \ion{Mg}{2} &  $+0.660$ &   \nodata & 11.5690 & 12.8559 & $^2D_{5/2}$          & $^2F^o_{7/2}$        & 1F & ID=25 \\
 9631.9470 & 9634.5889 & \ion{Mg}{2} &  $-0.640$ &   \nodata & 11.5690 & 12.8559 & $^2D_{5/2}$          & $^2F^o_{5/2}$        & 1F & ID=25 \\
 9632.4300 & 9635.0721 & \ion{Mg}{2} &  $+0.500$ &   \nodata & 11.5691 & 12.8559 & $^2D_{3/2}$          & $^2F^o_{5/2}$        & 1F & ID=25 \\
 9649.5710 & 9652.2177 & \ion{S}{1}  &  $+0.250$ &   \nodata &  8.4114 &  9.6960 & $^3D^o_{3}$          & $^3D_{3}$            & 2  & ID=26 \\
 9658.4343 & 9661.0834 & \ion{C}{1}  &  $-0.280$ &   \nodata &  7.4878 &  8.7711 & $^3P^o_{2}$          & $^3S_{1}$            & 1  & ID=27 \\
 9672.2840 & 9674.9369 & \ion{S}{1}  &  $-0.420$ &   \nodata &  8.4082 &  9.6897 & $^3D^o_{1}$          & $^3D_{1}$            & 3F & ID=28 \\
 9672.5320 & 9675.1849 & \ion{S}{1}  &  $-0.970$ &   \nodata &  8.4082 &  9.6896 & $^3D^o_{1}$          & $^3D_{2}$            & 3F & ID=28 \\
 9680.5610 & 9683.2161 & \ion{S}{1}  &  $-1.030$ &   \nodata &  8.4093 &  9.6897 & $^3D^o_{2}$          & $^3D_{1}$            & 2F & ID=29 \\
 9680.8090 & 9683.4642 & \ion{S}{1}  &  $-0.010$ &   \nodata &  8.4093 &  9.6896 & $^3D^o_{2}$          & $^3D_{2}$            & 2F & ID=29 \\
 9822.7500 & 9825.4436 & \ion{N}{1}  &  $-0.303$ &   \nodata & 11.7575 & 13.0194 & $^4D^o_{5/2}$        & $^4D_{5/2}$          & 3  & ID=30 \\
 9854.7588 & 9857.4611 & \ion{Ca}{2} &  $-0.205$ &   \nodata &  7.5051 &  8.7629 & $^2P^o_{1/2}$        & $^2S_{1/2}$          & 2  & ID=31 \\
 9863.3340 & 9866.0386 & \ion{N}{1}  &  $+0.080$ &   \nodata & 11.7638 & 13.0205 & $^4D^o_{7/2}$        & $^4D_{7/2}$          & 3  & ID=32 \\
 9890.6280 & 9893.3400 & \ion{Ca}{2} &  $+0.900$ &   \nodata &  8.4380 &  9.6912 & $^2F^o_{5/2}$        & $^2G_{7/2}$          & 1F & ID=33 \\
 9890.6280 & 9893.3400 & \ion{Ca}{2} &  $+1.013$ &   \nodata &  8.4380 &  9.6912 & $^2F^o_{7/2}$        & $^2G_{9/2}$          & 1F & ID=33 \\
 9890.6280 & 9893.3400 & \ion{Ca}{2} &  $-0.531$ &   \nodata &  8.4380 &  9.6912 & $^2F^o_{7/2}$        & $^2G_{7/2}$          & 1F & ID=33 \\
 9909.1100 & 9911.8271 & \ion{Fe}{2} &  $+1.160$ &   \nodata & 12.8225 & 14.0733 & 2[6]*                & 2[7]                 & 3B & ID=34 \\
 9909.7016 & 9912.4188 & \ion{S}{1}  &  $-0.768$ &   \nodata &  8.4093 &  9.6600 & $^3D^o_{2}$          & $^3P_{1}$            & 3B & ID=34 \\
 9910.0930 & 9912.8103 & \ion{Fe}{2} &  $+1.220$ &   \nodata & 12.8226 & 14.0733 & 2[6]*                & 2[7]                 & 3B & ID=34 \\
 9931.3741 & 9934.0972 & \ion{Ca}{2} &  $+0.092$ &   \nodata &  7.5148 &  8.7629 & $^2P^o_{3/2}$        & $^2S_{1/2}$          & 1  & ID=35 \\
 9947.0660 & 9949.7933 & \ion{N}{1}  &  $-1.140$ &   \nodata & 11.7575 & 13.0036 & $^4D^o_{5/2}$        & $^2F_{7/2}$          & 3B & ID=36 \\
 9947.8380 & 9950.5655 & \ion{Fe}{2} &  $+1.280$ &   \nodata & 12.7754 & 14.0214 & 2[7]*                & 2[8]                 & 3B & ID=36 \\
 9997.5980 & 10000.339 & \ion{Fe}{2} &  $-1.867$ &   \nodata &  5.4841 &  6.7239 & z\,$^4F^o_{9/2}$     & b\,$^4G_{1/2}$       & 2  & ID=37 \\
 10049.373 & 10052.128 & \ion{H}{1} (Pa\,$\delta$)  &  $-0.303$ & \nodata & 12.0875 & 13.3209 & $n=3$                & $n=7$ & 1 & ID=38 \\
 10092.095 & 10094.862 & \ion{Mg}{2} &  $+0.910$ &   $+0.96$ & 11.6297 & 12.8579 & $^2F^o_{5/2}$        & $^2G_{7/2}$          & 1F & ID=39 \\
 10092.217 & 10094.984 & \ion{Mg}{2} &  $+1.020$ &   $+1.07$ & 11.6297 & 12.8579 & $^2F^o_{7/2}$        & $^2G_{9/2}$          & 1F & ID=39 \\
 10092.217 & 10094.984 & \ion{Mg}{2} &  $-0.530$ &   $-0.48$ & 11.6297 & 12.8579 & $^2F^o_{7/2}$        & $^2G_{7/2}$          & 1F & ID=39 \\
 10105.132 & 10107.902 & \ion{N}{1}  &  $+0.235$ &   $+0.35$ & 11.7501 & 12.9767 & $^4D^o_{1/2}$        & $^4F_{3/2}$          & 2  & ID=40 \\
 10108.892 & 10111.663 & \ion{N}{1}  &  $+0.443$ &   \nodata & 11.7529 & 12.9790 & $^4D^o_{3/2}$        & $^4F_{5/2}$          & 2  & ID=41 \\
 10112.481 & 10115.253 & \ion{N}{1}  &  $+0.622$ &   $+0.59$ & 11.7575 & 12.9832 & $^4D^o_{5/2}$        & $^4F_{7/2}$          & 2  & ID=42 \\
 10114.640 & 10117.413 & \ion{N}{1}  &  $+0.777$ &   $+0.81$ & 11.7638 & 12.9893 & $^4D^o_{7/2}$        & $^4F_{9/2}$          & 1  & ID=43 \\
 10123.866 & 10126.642 & \ion{C}{1}  &  $-0.031$ &   $-0.09$ &  8.5371 &  9.7614 & $^1P_{1}$            & $^1P^o_{1}$          & 1  & ID=44 \\
 10147.267 & 10150.049 & \ion{N}{1}  &  $-0.169$ &   \nodata & 11.7575 & 12.9790 & $^4D^o_{5/2}$        & $^4F_{5/2}$          & 3  & ID=45 \\
 10164.848 & 10167.634 & \ion{N}{1}  &  $-0.323$ &   \nodata & 11.7638 & 12.9832 & $^4D^o_{7/2}$        & $^4F_{7/2}$          & 3  & ID=46 \\
 10173.515 & 10176.303 & \ion{Fe}{2} &  $-2.736$ &   $-2.79$ &  5.5107 &  6.7291 & z\,$^4D^o_{7/2}$     & b\,$^4G_{9/2}$       & 3  & ID=47 \\
 10216.313 & 10219.113 & \ion{Fe}{1} &  $-0.063$ &   $-0.29$ &  4.7331 &  5.9464 & y\,$^3D^o_{3}$       & e\,$^3F_{4}$         & 3  & ID=48 \\
 10245.556 & 10248.364 & \ion{Fe}{2} &  $-2.057$ &   $-1.98$ &  6.7303 &  7.9401 & b\,$^4G_{7/2}$       & y\,$^4G^o_{7/2}$     & 3  & ID=49 \\
 10327.311 & 10330.141 & \ion{Sr}{2} &  $-0.353$ &   $-0.40$ &  1.8395 &  3.0397 & $^2D_{5/2}$          & $^2P^o_{3/2}$        & 2  & ID=50 \\
 10332.928 & 10335.760 & \ion{Fe}{2} &  $-1.968$ &   \nodata &  6.7291 &  7.9286 & b\,$^4G_{9/2}$       & y\,$^4G^o_{9/2}$     & 3  & ID=51 \\
 10366.167 & 10369.008 & \ion{Fe}{2} &  $-1.825$ &   $-1.76$ &  6.7239 &  7.9197 & b\,$^4G_{1/2}$       & y\,$^4G^o_{1/2}$     & 3  & ID=52 \\
 10371.263 & 10374.106 & \ion{Si}{1} &  $-0.705$ &   $-0.80$ &  4.9296 &  6.1248 & $^3P^o_{1}$          & $^3S_{1}$            & 3  & ID=53 \\
 10452.819 & 10455.684 & \ion{C}{1}  &  $-0.722$ &   $-1.03$ &  9.6954 & 10.8813 & $^3F^o_{2}$          & 2[7/2]               & 3  & ID=54 \\
 10455.470 & 10458.335 & \ion{S}{1}  &  $+0.250$ &   $+0.33$ &  6.8601 &  8.0457 & $^3S^o_{1}$          & $^3P_{2}$            & 1  & ID=55 \\
 10456.790 & 10459.656 & \ion{S}{1}  &  $-0.447$ &   $-0.47$ &  6.8601 &  8.0455 & $^3S^o_{1}$          & $^3P_{0}$            & 2  & ID=56 \\
 10459.460 & 10462.326 & \ion{S}{1}  &  $+0.030$ &   $+0.08$ &  6.8601 &  8.0452 & $^3S^o_{1}$          & $^3P_{1}$            & 1  & ID=57 \\
 10463.006 & 10465.874 & \ion{Fe}{2} &  $-2.417$ &   $-2.33$ & 6.8031  &  7.9877 & d\,$^2F_{5/2}$       & z\,$^2F^o_{5/2}$     & 3  & ID=58 \\
 10470.970 & 10473.840 & \ion{C}{1}  &  $-0.672$ &   \nodata &  9.6975 & 10.8812 & $^3F^o_{3}$          & 2[7/2]               & 3  & ID=59 \\
 10501.503 & 10504.380 & \ion{Fe}{2} &  $-2.086$ &   $-2.17$ &  5.5488 &  6.7291 & z\,$^4F^o_{7/2}$     & b\,$^4G_{9/2}$       & 2  & ID=60 \\
 10507.000 & 10509.879 & \ion{N}{1}  &  $+0.118$ &   $+0.23$ & 11.8397 & 13.0194 & $^4P^o_{3/2}$        & $^4D_{5/2}$          & 3  & ID=61 \\
 10513.410 & 10516.291 & \ion{N}{1}  &  $-0.198$ &   $-0.10$ & 11.8374 & 13.0164 & $^4P^o_{1/2}$        & $^4D_{1/2}$          & 3  & ID=62 \\
 10520.580 & 10523.463 & \ion{N}{1}  &  $+0.024$ &   $+0.26$ & 11.8397 & 13.0179 & $^4P^o_{3/2}$        & $^4D_{3/2}$          & 3  & ID=63 \\
 10525.149 & 10528.034 & \ion{Fe}{2} &  $-2.958$ &   $-3.15$ &  5.5526 &  6.7303 & z\,$^4D^o_{5/2}$     & b\,$^4G_{7/2}$       & 3  & ID=64 \\
 10539.575 & 10542.463 & \ion{N}{1}  &  $+0.530$ &   $+0.60$ & 11.8445 & 13.0205 & $^4P^o_{5/2}$        & $^4D_{7/2}$          & 2  & ID=65 \\
 10541.227 & 10544.115 & \ion{C}{1}  &  $-1.398$ &   $-1.27$ &  8.5371 &  9.7130 & $^1P_{1}$            & $^1P^o_{1}$          & 3  & ID=66 \\
 10546.488 & 10549.378 & \ion{Fe}{2} &  $-0.303$ &   $+0.91$ &  9.6536 & 10.8289 & e\,$^6D_{9/2}$       & y\,$^6F^o_{1/2}$     & 3  & ID=67 \\
 10549.640 & 10552.531 & \ion{N}{1}  &  $+0.092$ &   $+0.15$ & 11.8445 & 13.0194 & $^4P^o_{5/2}$        & $^4D_{5/2}$          & 3  & ID=68 \\
 10585.141 & 10588.042 & \ion{Si}{1} &  $+0.012$ &   $-0.06$ &  4.9538 &  6.1248 & $^3P^o_{2}$          & $^3S_{1}$            & 2  & ID=69 \\
 10603.425 & 10606.330 & \ion{Si}{1} &  $-0.305$ &   $-0.37$ &  4.9296 &  6.0986 & $^3P^o_{1}$          & $^3P_{2}$            & 2  & ID=70 \\
 10635.970 & 10638.884 & \ion{S}{1}  &  $+0.460$ &   $+0.38$ &  8.5844 &  9.7501 & $^1D^o_{2}$          & $^1F_{3}$            & 2  & ID=71 \\
 10643.980 & 10646.896 & \ion{N}{1}  &  $-0.639$ &   \nodata & 11.8397 & 13.0042 & $^4P^o_{3/2}$        & $^4P_{1/2}$          & 3  & ID=72 \\
 10653.040 & 10655.959 & \ion{N}{1}  &  $-0.211$ &   \nodata & 11.8374 & 13.0009 & $^4P^o_{1/2}$        & $^4P_{3/2}$          & 3  & ID=73 \\
 10660.973 & 10663.893 & \ion{Si}{1} &  $-0.266$ &   $-0.32$ &  4.9201 &  6.0827 & $^3P^o_{0}$          & $^3P_{1}$            & 2  & ID=74 \\
 10675.668 & 10678.593 & \ion{O}{1}  &  $-0.351$ &   \nodata & 12.0786 & 13.2397 & $^5D^o_{4}$          & $^5F_{5}$            & 3F & ID=75 \\
 10675.668 & 10678.593 & \ion{O}{1}  &  $-1.216$ &   \nodata & 12.0786 & 13.2397 & $^5D^o_{4}$          & $^5F_{4}$            & 3F & ID=75 \\
 10675.668 & 10678.593 & \ion{O}{1}  &  $-2.392$ &   \nodata & 12.0786 & 13.2397 & $^5D^o_{4}$          & $^5F_{3}$            & 3F & ID=75 \\
 10675.766 & 10678.691 & \ion{O}{1}  &  $-0.516$ &   \nodata & 12.0786 & 13.2397 & $^5D^o_{3}$          & $^5F_{4}$            & 3F & ID=75 \\
 10675.766 & 10678.691 & \ion{O}{1}  &  $-1.070$ &   \nodata & 12.0786 & 13.2397 & $^5D^o_{3}$          & $^5F_{3}$            & 3F & ID=75 \\
 10675.766 & 10678.691 & \ion{O}{1}  &  $-2.091$ &   \nodata & 12.0786 & 13.2397 & $^5D^o_{3}$          & $^5F_{2}$            & 3F & ID=75 \\
 10675.906 & 10678.831 & \ion{O}{1}  &  $-0.710$ &   \nodata & 12.0786 & 13.2397 & $^5D^o_{2}$          & $^5F_{3}$            & 3F & ID=75 \\
 10675.906 & 10678.831 & \ion{O}{1}  &  $-0.710$ &   \nodata & 12.0786 & 13.2397 & $^5D^o_{2}$          & $^5F_{2}$            & 3F & ID=75 \\
 10675.906 & 10678.831 & \ion{O}{1}  &  $-1.091$ &   \nodata & 12.0786 & 13.2397 & $^5D^o_{2}$          & $^5F_{1}$            & 3F & ID=75 \\
 10676.023 & 10678.948 & \ion{O}{1}  &  $-0.944$ &   \nodata & 12.0787 & 13.2397 & $^5D^o_{1}$          & $^5F_{2}$            & 3F & ID=75 \\
 10676.023 & 10678.948 & \ion{O}{1}  &  $-1.245$ &   \nodata & 12.0787 & 13.2397 & $^5D^o_{1}$          & $^5F_{1}$            & 3F & ID=75 \\
 10676.079 & 10679.004 & \ion{O}{1}  &  $-1.245$ &   \nodata & 12.0787 & 13.2397 & $^5D^o_{0}$          & $^5F_{1}$            & 3F & ID=75 \\
 10683.080 & 10686.007 & \ion{C}{1}  &  $+0.079$ &   $+0.03$ &  7.4828 &  8.6430 & $^3P^o_{1}$          & $^3D_{2}$            & 1  & ID=76 \\
 10685.340 & 10688.268 & \ion{C}{1}  &  $-0.272$ &   $-0.30$ &  7.4804 &  8.6404 & $^3P^o_{0}$          & $^3D_{1}$            & 1  & ID=77 \\
 10691.245 & 10694.174 & \ion{C}{1}  &  $+0.344$ &   $+0.28$ &  7.4878 &  8.6472 & $^3P^o_{2}$          & $^3D_{3}$            & 1  & ID=78 \\
 10694.251 & 10697.181 & \ion{Si}{1} &  $+0.048$ &   $+0.10$ &  5.9639 &  7.1230 & $^3D_{2}$            & $^3F^o_{3}$          & 2  & ID=79 \\
 10707.320 & 10710.253 & \ion{C}{1}  &  $-0.411$ &   $-0.41$ &  7.4828 &  8.6404 & $^3P^o_{1}$          & $^3D_{1}$            & 1  & ID=80 \\
 10713.548 & 10716.483 & \ion{N}{1}  &  $-0.131$ &   \nodata & 11.8397 & 12.9967 & $^4P^o_{3/2}$        & $^4P_{5/2}$          & 3  & ID=81 \\
 10727.406 & 10730.345 & \ion{Si}{1} &  $+0.217$ &   $+0.29$ &  5.9840 &  7.1395 & $^3D_{3}$            & $^3F^o_{4}$          & 2  & ID=82 \\
 10729.529 & 10732.468 & \ion{C}{1}  &  $-0.420$ &   $-0.46$ &  7.4878 &  8.6430 & $^3P^o_{2}$          & $^3D_{2}$            & 1  & ID=83 \\
 10749.378 & 10752.323 & \ion{Si}{1} &  $-0.205$ &   $-0.21$ &  4.9296 &  6.0827 & $^3P^o_{1}$          & $^3P_{1}$            & 2  & ID=84 \\
 10753.980 & 10756.926 & \ion{C}{1}  &  $-1.606$ &   $-1.69$ &  7.4878 &  8.6404 & $^3P^o_{2}$          & $^3D_{1}$            & 2  & ID=85 \\
 10757.887 & 10760.834 & \ion{N}{1}  &  $-0.389$ &   $+0.05$ & 11.8445 & 12.9967 & $^4P^o_{5/2}$        & $^4P_{5/2}$          & 3  & ID=86 \\
 10786.849 & 10789.804 & \ion{Si}{1} &  $-0.303$ &   $-0.38$ &  4.9296 &  6.0787 & $^3P^o_{1}$          & $^3P_{0}$            & 3  & ID=87 \\
 10811.053 & 10814.015 & \ion{Mg}{1} &  $+0.024$ &   $+0.01$ &  5.9459 &  7.0924 & $^3D_{3}$            & $^3F^o_{4}$          & 2F & ID=88 \\
 10811.084 & 10814.046 & \ion{Mg}{1} &  $-0.137$ &   $-0.16$ &  5.9459 &  7.0924 & $^3D_{2}$            & $^3F^o_{3}$          & 2F & ID=88 \\
 10811.097 & 10814.059 & \ion{Mg}{1} &  $-1.038$ &   $-0.32$ &  5.9459 &  7.0924 & $^3D_{3}$            & $^3F^o_{3}$          & 2F & ID=88 \\
 10811.122 & 10814.084 & \ion{Mg}{1} &  $-1.036$ &   $-1.05$ &  5.9459 &  7.0924 & $^3D_{2}$            & $^3F^o_{2}$          & 2F & ID=88 \\
 10811.158 & 10814.120 & \ion{Mg}{1} &  $-0.305$ &   $-1.05$ &  5.9459 &  7.0924 & $^3D_{1}$            & $^3F^o_{2}$          & 2F & ID=88 \\
 10811.198 & 10814.160 & \ion{Mg}{1} &  $-0.190$ &   $-1.93$ &  5.9459 &  7.0924 & $^3D_{2}$            & $^1F^o_{3}$          & 2F & ID=88 \\
 10811.219 & 10814.181 & \ion{Mg}{1} &  $-1.280$ &   $-1.46$ &  5.9459 &  7.0924 & $^3D_{3}$            & $^1F^o_{3}$          & 2F & ID=88 \\
 10827.088 & 10830.054 & \ion{Si}{1} &  $+0.302$ &   $+0.23$ &  4.9538 &  6.0986 & $^3P^o_{2}$          & $^3P_{2}$            & 1  & ID=89 \\
 10843.858 & 10846.828 & \ion{Si}{1} &  $-0.112$ &   $-0.05$ &  5.8625 &  7.0055 & $^1P_{1}$            & $^1D^o_{2}$          & 3  & ID=90 \\
 10862.652 & 10865.628 & \ion{Fe}{2} &  $-2.199$ &   $-2.11$ &  5.5892 &  6.7303 & z\,$^4F^o_{5/2}$     & b\,$^4G_{7/2}$       & 2  & ID=91 \\
 10868.789 & 10871.767 & \ion{Si}{1} &  $+0.206$ &   $-0.01$ &  6.1910 &  7.3314 & $^3F^o_{3}$          & 2[9/2]               & 3  & ID=92 \\
 10869.536 & 10872.514 & \ion{Si}{1} &  $+0.371$ &   $+0.36$ &  5.0823 &  6.2227 & $^1P^o_{1}$          & $^1D_{2}$            & 2  & ID=93 \\
 10885.333 & 10888.314 & \ion{Si}{1} &  $+0.221$ &   $-0.10$ &  6.1807 &  7.3194 & $^3F^o_{2}$          & 2[7/2]               & 3  & ID=94 \\
 10914.244 & 10917.234 & \ion{Mg}{2} &  $+0.020$ &   $+0.00$ &  8.8637 &  9.9993 & $^2D_{5/2}$          & $^2P^o_{3/2}$        & 1B & ID=95 \\
 10914.887 & 10917.877 & \ion{Sr}{2} &  $-0.638$ &   $-0.59$ &  1.8047 &  2.9403 & $^2D_{3/2}$          & $^2P^o_{1/2}$        & 1B & ID=95 \\
 10915.284 & 10918.274 & \ion{Mg}{2} &  $-0.930$ &   $-1.00$ &  8.8638 &  9.9993 & $^2D_{3/2}$          & $^2P^o_{3/2}$        & 1B & ID=95 \\
 10938.093 & 10941.089 & \ion{H}{1} (Pa\,$\gamma$)  &  $+0.002$ & \nodata & 12.0875 & 13.2207 & $n=3$                & $n=6$        & 1 & ID=96 \\
 10951.778 & 10954.778 & \ion{Mg}{2} &  $-0.230$ &   $-0.33$ &  8.8638 &  9.9955 & $^2D_{3/2}$          & $^2P^o_{1/2}$        & 1  & ID=97 \\
 10965.450 & 10968.454 & \ion{Mg}{1} &  $-0.240$ &   $-1.15$ &  5.9328 &  7.0632 & $^3P^o_{2}$          & $^3D_{3}$            & 3  & ID=98 \\
 10979.308 & 10982.315 & \ion{Si}{1} &  $-0.524$ &   $-0.60$ &  4.9538 &  6.0827 & $^3P^o_{2}$          & $^3P_{1}$            & 3  & ID=99 \\
 10982.058 & 10985.066 & \ion{Si}{1} &  $+0.104$ &   $-0.27$ &  6.1910 &  7.3197 & $^3F^o_{3}$          & 2[7/2]               & 3  & ID=100 \\
 11017.966 & 11020.984 & \ion{Si}{1} &  $+0.760$ &   $+0.31$ &  6.2060 &  7.3310 & $^3F^o_{4}$          & 2[9/2]               & 2  & ID=101 \\
 11125.593 & 11128.640 & \ion{Fe}{2} &  $-2.300$ &   $-2.27$ &  5.6152 &  6.7293 & z\,$^4F^o_{3/2}$     & b\,$^4G_{5/2}$       & 3  & ID=102 \\
 11601.764 & 11604.940 & \ion{S}{1}  &  $-0.273$ &   $-0.15$ &  8.5844 &  9.6528 & $^1D^o_{2}$          & $^1P_{1}$            & 3  & ID=103 \\
 11619.282 & 11622.463 & \ion{C}{1}  &  $-0.574$ &   $-0.62$ &  8.6404 &  9.7072 & $^3D_{1}$            & $^3D^o_{1}$          & 2  & ID=104 \\
 11628.830 & 11632.014 & \ion{C}{1}  &  $-0.260$ &   $-0.39$ &  8.6430 &  9.7089 & $^3D_{2}$            & $^3D^o_{2}$          & 1  & ID=105 \\
 11647.977 & 11651.166 & \ion{C}{1}  &  $-1.016$ &   $-0.83$ &  8.6430 &  9.7072 & $^3D_{2}$            & $^3D^o_{1}$          & 3  & ID=106 \\
 11652.846 & 11656.036 & \ion{C}{1}  &  $-0.769$ &   $-0.87$ &  8.7711 &  9.8348 & $^3S_{1}$            & $^3P^o_{0}$          & 3  & ID=107 \\
 11658.820 & 11662.012 & \ion{C}{1}  &  $-0.278$ &   $-0.36$ &  8.7711 &  9.8343 & $^3S_{1}$            & $^3P^o_{1}$          & 1F & ID=108 \\
 11659.680 & 11662.872 & \ion{C}{1}  &  $+0.028$ &   $-0.07$ &  8.6472 &  9.7102 & $^3D_{3}$            & $^3D^o_{3}$          & 1F & ID=108 \\
 11669.626 & 11672.820 & \ion{C}{1}  &  $-0.030$ &   $-0.01$ &  8.7711 &  9.8333 & $^3S_{1}$            & $^3P^o_{2}$          & 1  & ID=109 \\
 11674.140 & 11677.336 & \ion{C}{1}  &  $-0.795$ &   $-0.90$ &  8.6472 &  9.7089 & $^3D_{3}$            & $^3D^o_{2}$          & 2  & ID=110 \\
 11748.220 & 11751.436 & \ion{C}{1}  &  $+0.375$ &   $+0.40$ &  8.6404 &  9.6954 & $^3D_{1}$            & $^3F^o_{2}$          & 1  & ID=111 \\
 11753.320 & 11756.537 & \ion{C}{1}  &  $+0.691$ &   $+0.69$ &  8.6472 &  9.7018 & $^3D_{3}$            & $^3F^o_{4}$          & 1  & ID=112 \\
 11754.760 & 11757.978 & \ion{C}{1}  &  $+0.542$ &   $+0.51$ &  8.6430 &  9.6975 & $^3D_{2}$            & $^3F^o_{3}$          & 1  & ID=113 \\
 11777.540 & 11780.764 & \ion{C}{1}  &  $-0.520$ &   $-0.59$ &  8.6430 &  9.6954 & $^3D_{2}$            & $^3F^o_{2}$          & 2  & ID=114 \\
 11801.080 & 11804.310 & \ion{C}{1}  &  $-0.735$ &   $-0.80$ &  8.6472 &  9.6975 & $^3D_{3}$            & $^3F^o_{3}$          & 2  & ID=115 \\
 11828.171 & 11831.409 & \ion{Mg}{1} &  $-0.333$ &   $-0.50$ &  4.3458 &  5.3937 & $^1P^o_{1}$          & $^1S_{0}$            & 1  & ID=116 \\
 11838.997 & 11842.238 & \ion{Ca}{2} &  $+0.312$ &   $+0.24$ &  6.4679 &  7.5148 & $^2S_{1/2}$          & $^2P^o_{3/2}$        & 1  & ID=117 \\
 11848.710 & 11851.953 & \ion{C}{1}  &  $-0.697$ &   $-0.70$ &  8.6430 &  9.6891 & $^3D_{2}$            & $^3P^o_{2}$          & 2  & ID=118 \\
 11862.985 & 11866.232 & \ion{C}{1}  &  $-0.710$ &   $-0.70$ &  8.6404 &  9.6852 & $^3D_{1}$            & $^3P^o_{1}$          & 2  & ID=119 \\
 11879.580 & 11882.832 & \ion{C}{1}  &  $-0.610$ &   $-0.65$ &  8.6404 &  9.6838 & $^3D_{1}$            & $^3P^o_{0}$          & 2  & ID=120 \\
 11892.898 & 11896.153 & \ion{C}{1}  &  $-0.277$ &   $-0.35$ &  8.6430 &  9.6852 & $^3D_{2}$            & $^3P^o_{1}$          & 1  & ID=121 \\
 11895.750 & 11899.006 & \ion{C}{1}  &  $-0.008$ &   $-0.02$ &  8.6472 &  9.6891 & $^3D_{3}$            & $^3P^o_{2}$          & 1  & ID=122 \\
 11949.744 & 11953.015 & \ion{Ca}{2} &  $+0.006$ &   $-0.04$ &  6.4679 &  7.5051 & $^2S_{1/2}$          & $^2P^o_{1/2}$        & 1  & ID=123 \\
 11984.198 & 11987.478 & \ion{Si}{1} &  $+0.239$ &   $+0.12$ &  4.9296 &  5.9639 & $^3P^o_{1}$          & $^3D_{2}$            & 3  & ID=124 \\
 11991.568 & 11994.850 & \ion{Si}{1} &  $-0.109$ &   $-0.22$ &  4.9201 &  5.9537 & $^3P^o_{0}$          & $^3D_{1}$            & 3  & ID=125 \\
 12031.504 & 12034.796 & \ion{Si}{1} &  $+0.477$ &   $+0.24$ &  4.9538 &  5.9840 & $^3P^o_{2}$          & $^3D_{3}$            & 2  & ID=126 \\
 12074.486 & 12077.791 & \ion{N}{1}  &  $-0.082$ &   \nodata & 12.0096 & 13.0362 & $^2D^o_{5/2}$        & $^2D_{5/2}$          & 3  & ID=127 \\
 12083.278 & 12086.585 & \ion{Mg}{1} &  $+0.450$ &   $-1.30$ &  5.7532 &  6.7790 & $^1D_{2}$            & $^3F^o_{3}$          & 2F & ID=128 \\
 12083.346 & 12086.653 & \ion{Mg}{1} &  $-0.790$ &   $-2.54$ &  5.7532 &  6.7790 & $^1D_{2}$            & $^3F^o_{2}$          & 2F & ID=128 \\
 12083.649 & 12086.956 & \ion{Mg}{1} &  $+0.410$ &   $+0.09$ &  5.7532 &  6.7790 & $^1D_{2}$            & $^1F^o_{3}$          & 2F & ID=128 \\
 12087.924 & 12091.232 & \ion{C}{1}  &  $-0.525$ &   $-0.77$ &  9.6954 & 10.7209 & $^3F^o_{2}$          & 2[7/2]               & 3  & ID=129 \\
 12103.534 & 12106.847 & \ion{Si}{1} &  $-0.350$ &   $-0.49$ &  4.9296 &  5.9537 & $^3P^o_{1}$          & $^3D_{1}$            & 3  & ID=130 \\
 12135.431 & 12138.752 & \ion{C}{1}  &  $+0.116$ &   \nodata &  9.7018 & 10.7231 & $^3F^o_{4}$          & 2[9/2]               & 2  & ID=131 \\
 12168.798 & 12172.129 & \ion{C}{1}  &  $-0.339$ &   $-0.40$ &  9.6954 & 10.7140 & $^3F^o_{2}$          & 2[7/2]               & 3  & ID=132 \\
 12186.840 & 12190.175 & \ion{N}{1}  &  $-0.005$ &   \nodata & 11.8445 & 12.8616 & $^4P^o_{5/2}$        & $^4P_{5/2}$          & 3  & ID=133 \\
 12192.945 & 12196.282 & \ion{C}{1}  &  $-0.258$ &   \nodata &  9.6975 & 10.7141 & $^3F^o_{3}$          & 2[7/2]               & 3  & ID=134 \\
 12244.357 & 12247.708 & \ion{C}{1}  &  $-2.008$ &   \nodata &  9.7018 & 10.7141 & $^3F^o_{4}$          & 2[7/2]               & 3B & ID=135 \\
 12244.683 & 12248.033 & \ion{C}{1}  &  $-1.959$ &   \nodata &  9.7102 & 10.7225 & $^3D^o_{3}$          & 2[5/2]               & 3B & ID=135 \\
 12244.875 & 12248.225 & \ion{C}{1}  &  $-0.784$ &   \nodata &  9.7102 & 10.7225 & $^3D^o_{3}$          & 2[5/2]               & 3B & ID=135 \\
 12248.696 & 12252.048 & \ion{C}{1}  &  $-0.677$ &   $-0.65$ &  9.7089 & 10.7209 & $^3D^o_{2}$          & 2[7/2]               & 3  & ID=136 \\
 12264.283 & 12267.639 & \ion{C}{1}  &  $-0.272$ &   $-0.25$ &  9.7102 & 10.7209 & $^3D^o_{3}$          & 2[7/2]               & 3  & ID=137 \\
 12270.692 & 12274.050 & \ion{Si}{1} &  $-0.396$ &   $-0.54$ &  4.9538 &  5.9639 & $^3P^o_{2}$          & $^3D_{2}$            & 3  & ID=138 \\
 12328.760 & 12332.134 & \ion{N}{1}  &  $+0.074$ &   \nodata & 11.9956 & 13.0009 & $^4S^o_{3/2}$        & $^4P_{3/2}$          & 3  & ID=139 \\
 12335.624 & 12338.999 & \ion{C}{1}  &  $-0.532$ &   $-0.61$ &  9.7089 & 10.7137 & $^3D^o_{2}$          & 2[5/2]               & 3  & ID=140 \\
 12461.253 & 12464.663 & \ion{N}{1}  &  $+0.405$ &   \nodata & 12.0001 & 12.9948 & $^2D^o_{3/2}$        & $^2F_{5/2}$          & 3  & ID=141 \\
 12463.840 & 12467.250 & \ion{O}{1}  &  $+0.104$ &   \nodata & 12.0786 & 13.0731 & $^5D^o_{4}$          & $^5F_{5}$            & 2F & ID=142 \\
 12463.840 & 12467.250 & \ion{O}{1}  &  $-0.761$ &   \nodata & 12.0786 & 13.0731 & $^5D^o_{4}$          & $^5F_{4}$            & 2F & ID=142 \\
 12463.840 & 12467.250 & \ion{O}{1}  &  $-1.937$ &   \nodata & 12.0786 & 13.0731 & $^5D^o_{4}$          & $^5F_{3}$            & 2F & ID=142 \\
 12463.974 & 12467.384 & \ion{O}{1}  &  $-0.061$ &   \nodata & 12.0786 & 13.0731 & $^5D^o_{3}$          & $^5F_{4}$            & 2F & ID=142 \\
 12463.974 & 12467.384 & \ion{O}{1}  &  $-0.614$ &   \nodata & 12.0786 & 13.0731 & $^5D^o_{3}$          & $^5F_{3}$            & 2F & ID=142 \\
 12463.974 & 12467.384 & \ion{O}{1}  &  $-1.637$ &   \nodata & 12.0786 & 13.0731 & $^5D^o_{3}$          & $^5F_{2}$            & 2F & ID=142 \\
 12464.165 & 12467.575 & \ion{O}{1}  &  $-0.255$ &   \nodata & 12.0786 & 13.0731 & $^5D^o_{2}$          & $^5F_{3}$            & 2F & ID=142 \\
 12464.165 & 12467.575 & \ion{O}{1}  &  $-0.637$ &   \nodata & 12.0786 & 13.0731 & $^5D^o_{2}$          & $^5F_{2}$            & 2F & ID=142 \\
 12464.165 & 12467.575 & \ion{O}{1}  &  $-1.636$ &   \nodata & 12.0786 & 13.0731 & $^5D^o_{2}$          & $^5F_{1}$            & 2F & ID=142 \\
 12464.325 & 12467.735 & \ion{O}{1}  &  $-0.490$ &   \nodata & 12.0787 & 13.0731 & $^5D^o_{1}$          & $^5F_{2}$            & 2F & ID=142 \\
 12464.325 & 12467.735 & \ion{O}{1}  &  $-0.790$ &   \nodata & 12.0787 & 13.0731 & $^5D^o_{1}$          & $^5F_{1}$            & 2F & ID=142 \\
 12464.401 & 12467.811 & \ion{O}{1}  &  $-0.790$ &   \nodata & 12.0787 & 13.0731 & $^5D^o_{0}$          & $^5F_{1}$            & 2F & ID=142 \\
 12469.615 & 12473.027 & \ion{N}{1}  &  $+0.610$ &   \nodata & 12.0096 & 13.0036 & $^2D^o_{5/2}$        & $^2F_{7/2}$          & 2  & ID=143 \\
 12549.479 & 12552.912 & \ion{C}{1}  &  $-0.565$ &   $-0.68$ &  8.8466 &  9.8343 & $^3P_{0}$            & $^3P^o_{1}$          & 2  & ID=144 \\
 12561.993 & 12565.430 & \ion{C}{1}  &  $-0.186$ &   \nodata &  9.7364 & 10.7231 & $^1F^o_{3}$          & 2[9/2]               & 2B & ID=145 \\
 12562.089 & 12565.526 & \ion{C}{1}  &  $-0.522$ &   $-0.65$ &  8.8481 &  9.8348 & $^3P_{1}$            & $^3P^o_{0}$          & 2B & ID=145 \\
 12569.032 & 12572.471 & \ion{C}{1}  &  $-0.598$ &   $-0.72$ &  8.8481 &  9.8343 & $^3P_{1}$            & $^3P^o_{1}$          & 2B & ID=146 \\
 12569.886 & 12573.325 & \ion{O}{1}  &  $-0.319$ &   \nodata & 12.0870 & 13.0731 & $^3D^o_{1}$          & $^3F_{2}$            & 2B & ID=146 \\
 12569.996 & 12573.435 & \ion{O}{1}  &  $-0.149$ &   \nodata & 12.0870 & 13.0731 & $^3D^o_{2}$          & $^3F_{3}$            & 2B & ID=146 \\
 12569.996 & 12573.435 & \ion{O}{1}  &  $-1.050$ &   \nodata & 12.0870 & 13.0731 & $^3D^o_{2}$          & $^3F_{2}$            & 2B & ID=146 \\
 12570.010 & 12573.449 & \ion{O}{1}  &  $-2.560$ &   \nodata & 12.0870 & 13.0731 & $^3D^o_{1}$          & $^3F_{1}$            & 2B & ID=146 \\
 12570.121 & 12573.560 & \ion{O}{1}  &  $-2.910$ &   \nodata & 12.0870 & 13.0731 & $^3D^o_{2}$          & $^3F_{1}$            & 2B & ID=146 \\
 12570.138 & 12573.577 & \ion{O}{1}  &  $+0.012$ &   \nodata & 12.0870 & 13.0731 & $^3D^o_{3}$          & $^3F_{4}$            & 2B & ID=146 \\
 12570.138 & 12573.577 & \ion{O}{1}  &  $-1.050$ &   \nodata & 12.0870 & 13.0731 & $^3D^o_{3}$          & $^3F_{3}$            & 2B & ID=146 \\
 12570.138 & 12573.577 & \ion{O}{1}  &  $-2.600$ &   \nodata & 12.0870 & 13.0731 & $^3D^o_{3}$          & $^3F_{2}$            & 2B & ID=146 \\
 12581.590 & 12585.032 & \ion{C}{1}  &  $-0.536$ &   $-0.67$ &  8.8481 &  9.8333 & $^3P_{1}$            & $^3P^o_{2}$          & 2  & ID=147 \\
 12590.812 & 12594.257 & \ion{C}{1}  &  $-0.631$ &   \nodata &  9.7364 & 10.7209 & $^1F^o_{3}$          & 2[7/2]               & 3  & ID=148 \\
 12601.466 & 12604.914 & \ion{C}{1}  &  $-0.443$ &   $-0.58$ &  8.8507 &  9.8343 & $^3P_{2}$            & $^3P^o_{1}$          & 2  & ID=149 \\
 12614.091 & 12617.542 & \ion{C}{1}  &  $+0.049$ &   $-0.06$ &  8.8507 &  9.8333 & $^3P_{2}$            & $^3P^o_{2}$          & 1  & ID=150 \\
 12818.077 & 12821.584 & \ion{H}{1} (Pa\,$\beta$)  &  $+0.433$ & \nodata & 12.0875 & 13.0545 & $n=3$                & $n=5$          & 1 & ID=151 \\
 13123.410 & 13126.999 & \ion{Al}{1} &  $+0.270$ &   $+0.11$ &  3.1427 &  4.0872 & $^2S_{1/2}$          & $^2P^o_{3/2}$        & 3  & ID=152 \\
 13150.753 & 13154.350 & \ion{Al}{1} &  $-0.030$ &   $-0.19$ &  3.1427 &  4.0853 & $^2S_{1/2}$          & $^2P^o_{1/2}$        & 3  & ID=153 \\
 13163.889 & 13167.489 & \ion{O}{1}  &  $-0.254$ &   $-0.33$ & 10.9888 & 11.9304 & $^3P_{1}$            & $^3S^o_{1}$          & 1F & ID=154 \\
 13164.858 & 13168.459 & \ion{O}{1}  &  $-0.032$ &   $-0.11$ & 10.9889 & 11.9304 & $^3P_{2}$            & $^3S^o_{1}$          & 1F & ID=154 \\
 13165.131 & 13168.732 & \ion{O}{1}  &  $-0.731$ &   $-0.80$ & 10.9889 & 11.9304 & $^3P_{0}$            & $^3S^o_{1}$          & 1F & ID=154 \\
 13176.888 & 13180.492 & \ion{Si}{1} &  $-0.200$ &   $-0.30$ &  5.8625 &  6.8031 & $^1P_{1}$            & $^1P^o_{1}$          & 2  & ID=155 \\
\enddata
\end{deluxetable*}

\begin{deluxetable*}{lll}
 \tablecaption{Unidentified features in 21 Lyn \label{tab:unID}}
 \tablehead{
 \multicolumn{2}{c}{Wavelength (\AA)} & \colhead{Notes} \\
 \cmidrule(r){1-2}
 \colhead{Air} & \colhead{Vacuum} & 
 }
 \startdata
 9121.1  & 9123.6  & Possibly a stellar line.  \ion{Cl}{1} $\lambda9121.14$? \\
 9258.3  & 9260.8  & Probably noise caused in normalization. \\
 9625.5  & 9628.1  & Possibly a stellar line, but seriously blended with telluric absorption. \\
 9634.2  & 9636.8  & Perhaps noise due to night-sky emission lines. \\
 9686.2  & 9688.9  & Perhaps a stellar line.  \ion{S}{1} $\lambda9685.87$? \\
 9701.3  & 9704.0  & Possibly a stellar line, but blended with telluric absorption. \\
 9811.5  & 9814.2  & Possibly a stellar line, but showing an asymmetric profile.  \ion{Fe}{2} $\lambda9811.34$? \\
 9830.2  & 9832.9  & Possibly a stellar line, but showing an asymmetric profile.  \ion{Fe}{2} $\lambda9830.50$? \\
 9858.9  & 9861.6  & Perhaps a stellar line.  \ion{C}{1} $\lambda9859.16$? \\
 9880.0  & 9882.7  & Likely noise because the dispersion of frames is relatively large at this part. \\
 10070.0 & 10072.8 & Probably noise due to an instrumental defect. \\
 10434.2 & 10437.1 & Likely a stellar line. \\
 10689.5 & 10692.4 & Likely a stellar line.  \ion{Si}{1} $\lambda10689.72$? \\
 10734.4 & 10737.3 & Likely noise because the dispersion of frames is relatively large at this part. \\
 10747.2 & 10750.1 & Likely noise. \\
 10831.4 & 10834.4 & Likely noise due to night-sky emission lines. \\
 11611.0 & 11614.2 & Likely noise due to heavy telluric absorption. \\
 11645.5 & 11648.7 & Probably noise because the dispersion of frames is large at this part. \\
 11790.7 & 11793.9 & Possibly a stellar line, but seriously blended with telluric absorption. \\
 11844.2 & 11847.4 & Likely noise due to an instrumental defect. \\
 11866.3 & 11869.5 & Probably noise because the dispersion of frames is large at this part. \\
 11867.8 & 11871.0 & Probably noise because the dispersion of frames is large at this part. \\
 11901.4 & 11904.7 & Probably noise due to an instrumental defect. \\
 12004.3 & 12007.6 & Probably noise because the dispersion of frames is large at this part. \\
 12172.0 & 12175.3 & Possibly a stellar line. \\
 12347.5 & 12350.9 & Possibly a stellar line.  \ion{C}{1} $\lambda12347.7$? \\
 12529.0 & 12532.4 & Probably noise due to an instrumental defect. \\
 12623.2 & 12626.7 & Probably noise because the dispersion of frames is large at this part. \\
 13241.6 & 13245.2 & Possibly a stellar line. \\
 \enddata
\end{deluxetable*}

\clearpage

%%%%%%%%%%%%%%%%%%%%%
% reference
%%%%%%%%%%%%%%%%%%%%%

\bibliographystyle{apj}

\end{document}